\documentclass[12pt]{article}
\pdfoutput=1
\usepackage{titling}
\usepackage{amsmath}
\usepackage{slashed}
\usepackage{amssymb}
\usepackage{epsfig}
\usepackage{graphicx}
\usepackage{multirow}
\usepackage{color}
\usepackage[normal]{subfigure}
\usepackage{rotating}
\usepackage{hyperref}
\usepackage[margin=0.9in]{geometry}
\usepackage[table]{xcolor}
\usepackage{enumitem}
\usepackage[utf8x]{inputenc}
\usepackage[compress,numbers,sort]{natbib}
\usepackage{authblk}
\usepackage{colortbl}
\usepackage{pdflscape}
\usepackage{color}
\usepackage{mathtools}
\usepackage{tabu}
\usepackage{braket}
\usepackage{bm}
\usepackage{booktabs}

\def\eq#1{{Eq.~(\ref{#1})}}
\def\eqs#1#2{{Eqs.~(\ref{#1})--(\ref{#2})}}
\def\fig#1{{Fig.~\ref{#1}}}

\def\Table#1{{Table~\ref{#1}}}

\def\sec#1{{Sec.~\ref{#1}}}

\def\app#1{{Appendix~\ref{#1}}}
\def\apps#1#2{{Appendices~\ref{#1} and \ref{#2}}}

\def\Re{\mbox{Re}}

\newcommand{\be} {\begin{equation}}
\newcommand{\ee} {\end{equation}}
\newcommand{\ba} {\begin{eqnarray}}
\newcommand{\ea} {\end{eqnarray}}

\newcommand{\published}[1]{%
\gdef\puB{#1}}
\newcommand{\puB}{}

\numberwithin{equation}{section}

\begin{document} 
\title{
Effective Theory Approach to New Physics with Flavour:\\ \Large
General Framework and a Leptoquark Example}

\author[ ]{\large Marzia Bordone\thanks{marzia.bordone@uni-siegen.de}}
\author[ ]{\large Oscar Cat\`a\thanks{oscar.cata@uni-siegen.de}}
\author[ ]{\large Thorsten Feldmann\thanks{thorsten.feldmann@uni-siegen.de}}

\affil[ ]{\emph{\normalsize Theoretische Physik 1, Naturwissenschaftlich-Technische Fakult\"at, Universit\"at Siegen\\ Walter-Flex-Stra\ss{}e 3,  D-57068 Siegen, Germany}}

\published{\flushright SI-HEP-2019-13, QFET-2019-10, P3H-19-032\vskip2cm}

\allowdisplaybreaks

\maketitle

\begin{abstract}
\noindent

Extending the Standard Model with higher-dimensional operators in an effective field theory (EFT) 
approach provides a systematic framework to study new physics (NP) effects from a bottom-up perspective, 
as long as the NP scale is sufficiently large compared to the energies probed in the experimental observables.
However, when taking into account the different quark and lepton flavours, the number of free parameters 
increases dramatically, which makes generic studies of the NP flavour structure  infeasible.
In this paper, we address this issue in view of the recently observed ``flavour anomalies'' in $B$-meson decays,
which we take as a motivation to develop a general framework that allows us to systematically reduce the 
number of flavour parameters in the EFT.  This framework can be easily used in global fits to 
flavour observables at Belle II and LHCb as well as in analyses of flavour-dependent collider signatures at
the LHC. Our formalism represents an extension 
of the well-known minimal-flavour-violation approach, and uses Froggatt-Nielsen charges to define 
the flavour power-counting.
As a relevant illustration of the formalism, we apply it to the flavour structures which could be induced by a  
$U_1$ vector leptoquark, which represents one of the possible  explanations for the 
recent hints of flavour non-universality in semileptonic $B$-decays. We study the phenomenological viability of this specific framework performing a fit to low-energy flavour observables.
 
\end{abstract}

\clearpage
\tableofcontents
\clearpage
\section{Introduction}
\label{sec:1}

The discovery of the Higgs boson and the scrutiny of physics close to the TeV scale --
so far without direct new physics (NP) detection -- provide strong evidence that the Standard Model (SM) is the right description of the fundamental interactions at the energies probed so far in collider-based experiments. 
Despite this, there are strong indications that the SM cannot be complete. In order to explore physics beyond the SM it is convenient to regard the SM Lagrangian as the leading term of an effective field theory (EFT) expansion, with corrections suppressed by inverse powers of the NP scale(s)
\cite{Buchmuller:1985jz,Grzadkowski:2010es}. The main advantage of such an EFT description is that it provides a general and systematic parametrisation of NP effects and thus enables a thorough scanning of the possible deviations from the SM. If NP is assumed to be weakly coupled, the relevant EFT is commonly referred to as Standard Model Effective Field Theory (SMEFT).

The origin of the different quark and lepton flavour species and an understanding of the hierarchical structure of flavour masses and couplings is one of the long-standing problems onto which NP should shed light. The SM provides the most general parametrisation of the flavour sector compatible with gauge symmetry, which turns out to be phenomenologically very successful. However, this requirement is not very constraining and the Yukawa sector turns out to be responsible for most of the parameters of the SM. The situation gets considerably worse when one works with SMEFT. Consider, as a concrete example, the EFT operator 
\begin{align}
\frac{1}{\Lambda^2}[{\mathcal{C}}_{lq}]^{ij\alpha\beta}({\bar{Q}}_i\gamma_{\mu}Q_j)({\bar{L}}_{\alpha}\gamma^{\mu}L_{\beta})\,,
\end{align}
where $Q$ and $L$ are left-handed quark and lepton fields, respectively, $i,j=1\ldots 3$ and $\alpha,\beta=1 \ldots 3$ their respective generation indices and $\Lambda$ is a generic NP scale. The flavour structure is encoded in the complex tensor coefficient $[{\mathcal{C}}_{lq}]^{ij\alpha\beta}$, which in a generic setup consists of 81 independent real entries. An attempt to constrain the flavour structure at NLO is therefore not realistic and additional assumptions in the direction of a theory of flavour are necessary.

Given our experience with the SM Yukawa couplings, the two main questions that we want to address are:
\begin{itemize}
 \item[(i)] What is the (relative) size of the individual entries in that tensor?
 \item[(ii)] Are there any (approximate) relations between the entries that reduce the 
    number of relevant parameters?
\end{itemize}
Answers to both questions can and should be provided  from a theoretical and phenomenological perspective. Regarding the latter, recently there have been indications that semileptonic $B$-meson decays might violate lepton flavour universality (LFU). Tensions have been observed in both $b\to s\ell\ell$ \cite{Aaij:2014ora,Aaij:2017vbb,Aaij:2019wad,Aaij:2015oid}  and $b\to c\ell\nu$ \cite{Lees:2012xj,Lees:2013uzd,Aaij:2015yra,Hirose:2016wfn,Hirose:2017dxl,Aaij:2017uff,Aaij:2017deq,Abdesselam:2019dgh} decays. At present, these are the strongest indications of non-trivial flavour structure beyond the SM. In this paper we will consider these indications as a guidance to answer the previous questions. 

Without further assumptions, the generic EFT power counting for dimensionless coupling constants would simply lead to $[{\mathcal{C}}_{lq}]^{ij\alpha\beta} \sim {\cal O}(1)$ with no priors on further flavour hierarchies or patterns. However, when confronted with precision flavour data, in particular for quark transitions between the second and first generations, this would require very large values of the NP scale $\Lambda$, way above the current LHC reach for direct detection. In particular, such high values of $\Lambda$ would not be able to explain the currently observed  ``anomalies'' in $B$-meson decays. 

One can adopt additional theoretical assumptions, with increasing levels of sophistication, in order to (a) set a phenomenologically acceptable power counting and (b) achieve a substantial reduction of the flavour parameters. 

A possibility would be to assume minimal flavour violation (MFV) \cite{DAmbrosio:2002vsn,Buras:2003jf}, where the flavour structure of $[{\mathcal{C}}_{lq}]^{ij\alpha\beta} $ is related to the SM Yukawa matrices. Technically, this is achieved by treating the SM Yukawa matrices as spurions which transform under the (broken) flavour symmetries in the quark and lepton sector, respectively. In our case this leads to 
 \begin{align}
\mbox{MFV:} \qquad &   [{\mathcal{C}}_{lq}]^{ij\alpha\beta} =
\left(\#  \delta^{ij} + \# (Y_U Y_U^\dagger)^{ij} + \#  (Y_D Y_D^\dagger)^{ij} + \ldots \right) 
\left( \delta^{\alpha\beta} + \ldots \right)\,,
 \end{align}
where $\#$ stands for an ${\mathcal O}(1)$ flavour-universal coefficient, and the dots indicate higher-order terms with additional suppressions by fermion masses and mixing angles. Here we did not include any new sources of flavour symmetry breaking in the lepton sector. By dropping the higher-order terms, the number of relevant NP parameters (six, for the above approximation) is drastically reduced compared to the generic case. In the context of MFV, the NP scale could be as low as a few TeV, which matches the expectations to accommodate measurable deviations from the SM in $B$-meson decays. However, MFV cannot account for the present indications of non-universal lepton-flavour couplings since, as in the SM, they are highly suppressed by the lepton Yukawa couplings.

A more general treatment of flavour is clearly required. A possibility is to introduce models with a full-fledged flavour structure, possibly considering particular mechanisms to break the flavour symmetries. Representative examples can be found e.g.\ in \cite{Barbieri:2011ci,Varzielas:2015iva,Hiller:2016kry,deMedeirosVarzielas:2019lgb}. 

The alternative is to provide an EFT-oriented setting, more in line with the MFV spirit, by extending the number of spurions. A step in this direction was already taken in \cite{Feldmann:2006jk}, where only  spurions 
related to quark bilinears were considered. The obvious challenges of this approach are
(a) which spurions to consider and (b) which flavour power counting should they obey. 

An interesting possibility to set a power counting is to adopt and generalise a suggestion by Froggatt and Nielsen \cite{Froggatt:1978nt}. They showed that by postulating a (spontaneously broken) new $U(1)$ symmetry with generation-dependent $U(1)$ charge assignments to each quark multiplet, one can fit remarkably well the SM flavour hierarchies, provided that a sufficient number of heavy fermions exist and that spontaneous breaking be triggered by the vacuum expectation value $\langle \phi_{\rm FN}\rangle $ of a new scalar field, which happens at a very high scale $\Lambda_{\rm FN} \gg \Lambda$. It is rather straightforward to generalise the model and include leptons. In the Froggatt-Nielsen (FN) model, flavour non-diagonal transitions are suppressed by powers of $\lambda=(\langle \phi_{\rm FN}\rangle/\Lambda_{\rm FN}) \ll 1$ and entirely specified by the corresponding charge differences. 

The FN model is a theory of flavour that generates the SM flavour hierarchies. In order to generate non-universal lepton-flavour structures in $[{\mathcal{C}}_{lq}]^{ij\alpha\beta}$ the model should be modified. 
In the following, we pursue a simpler approach and merely use the FN charges as a recipe for flavour structure, independently of a dynamical mechanism like Froggat and Nielsen advocated.\footnote{Actually, in a recent paper
\cite{Smolkovic:2019jow}
an {\it inverted} FN mechanism was proposed, where the expansion parameter is $m/\langle \phi\rangle$, $m$ being a mass parameter for new vector-like fermions.} 
If one adopts a generalised FN scenario  -- where both the quark and lepton sectors are charged -- as a power-counting scheme, one can 
extend it straightforwardly to SMEFT operators. In our example, 
\begin{align}
\mbox{FN:} \qquad &   [{\mathcal{C}}_{lq}]^{ij\alpha\beta} \sim 
\lambda^{\left| b_Q^i-b_Q^j+b_L^\alpha-b_L^\beta \right|} \,,
\end{align}
where $b_X^i$ are integer FN charges, and $\lambda \sim 0.2$ is usually associated with the Cabibbo angle. In this paper we will follow this approach and thus will not intend to build a full-fledged theory of flavour. A FN power-counting scheme for flavour hierarchies beyond the SM is appealing in different ways: first, it provides a reasonable phenomenological description of the SM flavour structure with a rather simple setup; and second, the power counting is automatically self-consistent, in a way to be explained in the next sections. An important consequence of such a scheme is that the relative size of different quark and lepton transitions is correlated. However, there is no reduction of parameters. 

In order to extend MFV it still remains to decide which new flavour structures, or spurions, to add. Taking the presently observed flavour anomalies at face value, possible NP explanations involve the exchange of relatively light new particles.

Among these, the simplest viable candidate is a vector leptoquark (LQ) usually dubbed $U_1$ in the literature. This leptoquark couples to the currents $(\bar Q \gamma_\mu L)$ and $(\bar d_R \gamma_\mu e_R)$. In such a situation, we can assume that the dominant NP effects are taken into account by allowing 
for two new fundamental flavour structures, which can be systematically implemented through two spurions, along the lines discussed in \cite{Feldmann:2006jk}. 
Back to our example, in a $U_1$ leptoquark model, the flavour structure factorises as 
\begin{align}
\mbox{LQ+FN:} \qquad &   [{\mathcal{C}}_{lq}]^{ij\alpha\beta} \sim
(\Delta_{QL})^{i\beta} \, (\Delta_{QL}^\dagger)^{\alpha j} + \ldots \sim 
\lambda^{\left| b_Q^i-b_L^\beta \right|} \,
\lambda^{\left| b_L^\alpha-b_Q^j \right|} \,,
\end{align}
and is dictated by a single spurion $(\Delta_{QL})^{i\beta}$.
The FN power counting dictates its hierarchical structure, which is determined by the FN charges. 
Notice that, as a consequence of triangle inequalities, 
\begin{align}
\left| b_Q^i-b_L^\beta \right| + \left| b_L^\alpha-b_Q^j \right| \geq \left| b_Q^i-b_Q^j+b_L^\alpha-b_L^\beta \right| \,,
\end{align}
the individual flavour coefficients in this scenario are always smaller or equal to the ones
in the unconstrained FN scenario above. This is one example of the kind of consistency conditions 
that one would have to require in an extended MFV approach, as emphasised in \cite{Feldmann:2006jk}. This extended MFV approach, with a new flavour spurion $\Delta_{QL}$ associated with
a fundamental leptoquark coupling, reduces the number of (relevant) independent parameters 
in $[{\mathcal{C}}_{lq}]^{ij\alpha\beta}$ to 9 complex entries plus one overall 
complex coefficient. 

The different kinds of theoretical assumptions discussed above together with the resulting number of independent parameters are summarised in \Table{tab:warmup}. 
\begin{table}[t!]
\def\arraystretch{1.5}
\begin{center} 
 \begin{tabular}{l c c}
  Approach & $[{\mathcal{C}}_{lq}]^{ij\alpha\beta}$  & NP parameters 
  \\
  \hline 
  generic EFT &  $ \sim {\mathcal O}(1)$ & 81
  \\
  MFV & $= \left( \# \delta^{ij} + \# (Y_U Y_U^\dagger)^{ij} + \#  (Y_D Y_D^\dagger)^{ij} \right) \delta^{\alpha\beta}$ & 6
  \\
  generic FN &  $\sim\apps{app:A}{app:2} 
\lambda^{\left| b_Q^i-b_Q^j+b_L^\alpha-b_L^\beta \right|}$ & 81
\\
LQ+FN &
$
= \# \, (\Delta_{QL})^{i\beta} \, (\Delta_{QL}^\dagger)^{\alpha j} 
\sim 
\lambda^{\left| b_Q^i-b_L^\beta \right|} \,
\lambda^{\left| b_L^\alpha-b_Q^j \right|} $ & 18 + 2
\end{tabular}
\end{center}
\caption{\label{tab:warmup} Cabibbo-scaling and factorisation of the flavour coefficient $\mathcal{C}_{lq}$
in different theoretical approaches, and the resulting reduction of NP parameters.}
\end{table}

In the rest of this paper we put these ideas on a general and more systematic ground. As an illustration of the method, we consider in detail the minimally extended MFV scenario that accounts for the present flavour anomalies. Since one of the most promising setups to fit the anomalies is a $U_1$ leptoquark model \cite{DiLuzio:2017vat,Calibbi:2017qbu,Barbieri:2017tuq,Blanke:2018sro,DiLuzio:2018zxy,Faber:2018qon,Heeck:2018ntp,Angelescu:2018tyl,Schmaltz:2018nls,Greljo:2018tzh,Fornal:2018dqn,Baker:2019sli,Cornella:2019hct,DaRold:2019fiw,Bordone:2017bld,Bordone:2018nbg,Buttazzo:2017ixm}, one simply needs to consider the spurions which are associated with the fermion currents that the $U_1$ leptoquark is coupling to (see below). These spurions will be referred to as $\Delta_{QL}$ and $\Delta_{DE}$ in 
the following. The power counting is dictated by the FN charges, which are partly constrained by the CKM matrix entries and the charged lepton masses\footnote{See \cite{Varzielas:2015iva} for previous attempts to combine FN charges with a $U_1$ vector leptoquark model.}. Other combinations which are not constrained in the SM will be fixed from phenomenological requirements on different flavour physics observables. This reduces the number of solutions (i.e.\ FN charge assignments that comply with phenomenology) to 11.   

We examine the viability of these 11 solutions by performing a fit to relevant low-energy observables. It turns out that all the solutions have a comparable figure of merit. However, they can be distinguished if more precise measurements are performed on e.g.\ $\bar B_{s,d}\to \tau^{\pm}\mu^{\mp}$. One should emphasise that when leptoquarks are involved, processes are not necessarily symmetric when the lepton charges are flipped.
In particular, we find substantially larger predictions for the $\bar B_{s,d}\to \tau^-\mu^+$ modes compared 
to $\bar B_{s,d} \to \tau^+\mu^-$. 

This paper is organised as follows: in \sec{sec:2} we catalog the spurions that break flavour symmetry assuming SM particle content. In the same section we introduce the FN power counting for quarks and leptons 
and summarise the existing constraints on the FN charges within the SM.
In \sec{sec:U1_model} we consider a simplified scenario where the spurions that act as an additional source of flavour-symmetry breaking are associated with the exchange of  a vector leptoquark $U_1$. 
We work out the resulting constraints on the FN charges following from various phenomenological 
considerations. In \sec{sec:fit} we further explore the phenomenological viability of our approach,
by performing a fit to low-energy flavour data, and discuss the results for a number of different 
scenarios that are distinguished by different FN charge assignments.
Concluding remarks are given in \sec{sec:conclusions}. Formulae for the observables included in the fit are provided in \apps{app:A}{app:2}. In \app{app:scaling} we collect the FN power counting for the flavour spurions and rotation matrices for the scenarios identified through the fit.

\section{Flavour Structure of New Physics Operators}
\label{sec:2}

\subsection{General spurion analysis of flavour}
\label{sec:spurions}

The starting point is the maximal flavour symmetry group of the SM commuting with the gauge symmetries, namely 
\begin{align}
{\cal G}_f=SU(3)_Q\times SU(3)_U\times SU(3)_D\times SU(3)_L\times SU(3)_E
\,. 
\end{align}
Here we only take into account rotations in generation space. Additional $U(1)$ factors are not essential for the purpose of our paper and will be ignored in the following. 

In the SM, the flavour symmetry ${\cal{G}}_f$ is broken by the Yukawa couplings of fermions to the Higgs field,\footnote{In the minimal SM the $U(1)$ symmetries describing lepton flavour conservation remain unbroken.}
\begin{align}
\label{eq:Yukawa}
{\cal{L}}_Y=
- {\bar{Q}} \, Y_U \, \tilde\varphi \, u 
- {\bar{Q}} \, Y_D \, \varphi \, d - {\bar{L}} \, Y_E \varphi \,  e + \mbox{h.c.}\,,
\end{align}
where $\varphi$ is the Higgs field and $\tilde{\varphi}$ its charge conjugate. $Q$ and $L$ are the weak left-handed quark and lepton doublets, and $u,d,e$ the right-handed weak singlets. In the following, the chirality of these 
fields is implicitly understood and the corresponding indices are not shown for simplicity.

The elements of the Yukawa matrices can be considered as perturbations, except for the element $(Y_U)_{33}$
which in the SM turns out to be ${\cal O}(1)$ and gives the top-quark a mass which is of the same order as the 
vacuum expectation value of the Higgs. In order to systematically incorporate the sources of flavour breaking, it is convenient to use a spurion analysis, where the Yukawa matrices are promoted to objects with definite transformations under the flavour symmetry, such that \eq{eq:Yukawa} is formally invariant under ${\cal G}_f$.     

In physics beyond the SM, flavour structures do not need to be restricted to the SM Yukawa matrices. In this Section we will consider new flavour structures which may be associated with the exchange of new, relatively heavy, bosonic particles. This subset of structures covers typical models with new scalars or vector bosons coupling to fermion bilinears. 
In this work, we restrict ourselves to spurions that couple to fermion bilinears ${\bar{\psi}}_i\Gamma \psi_j$, 
where $\psi_i$ are Dirac fields referring to SM quarks and leptons. We exclude right-handed Dirac and Majorana neutrinos from our analysis. 

For the classification of the possible structures, what matters are the involved fermion multiplets and 
their chirality, such that we only have to distinguish between scalar-like couplings 
\begin{align}
S&=\left\{S_0; i\gamma^5S_5; \sigma_{\mu\nu}S^{\mu\nu}\right\}\,,
\end{align}
which may be associated to the exchange of  scalar, pseudoscalar or tensor particles,
and couple left- and right-handed fields, 
or vector-like couplings 
\begin{align}
\Delta^{\mu}&=\left\{\Delta_V^{\mu}; \gamma^5\Delta_A^{\mu}\right\}\,,
\end{align} 
which may be associated with the exchange of vector or axial-vector particles and 
may induce left-left and right-right couplings.

The complete list of spurions comprises 30 structures, which, together with their quantum numbers under gauge and flavour symmetries, are collected in \Table{tab:1}. We list the spurions coupling to quark, lepton and mixed (leptoquark) bilinears, separated by double lines. Single lines separate the spurions which come along 
with currents carrying trivial and nontrivial baryon and/or lepton number, respectively.
\begin{table}[hbt!]
\begin{center}
\centering
\renewcommand{\arraystretch}{1.2} 
\resizebox{15cm}{!}{
\begin{tabular}{cc|cc|c}
Dirac bilinear &  $SU(3)\times SU(2)\times U(1)$ & Flavour spurion & $\mathcal{G}_f$ & $(\Delta B; \Delta L)$ \\
\hline
$\bar{Q}\gamma^\mu Q$ &  $(1\oplus 8,1\oplus 3,0)$ & $\Delta_Q$ & $(1\oplus 8,1,1)(1,1)$ & $\left(0;0\right)$ \\[1mm]
$\bar{u} \gamma^\mu u$ &  $(1\oplus 8,1,0)$ &$\Delta_U$ & $(1,1\oplus 8,1)(1,1)$ & $\left(0;0\right)$\\[1mm]
$\bar{d} \gamma^\mu d$ &  $(1\oplus 8,1,0)$ &$\Delta_D$ & $(1,1,1\oplus 8)(1,1)$ & $\left(0;0\right)$\\[1mm]
$\bar{u}\gamma^\mu d$ &  $(1\oplus 8,1,1)$ &$\Delta_{UD}$ & $(1,3,\bar{3})(1,1)$ & $\left(0;0\right)$\\[1mm]
$\bar{Q} u$ &  $(1\oplus 8,2,-\tfrac{1}{2})$ & $Y_U $ &$(3,\bar{3},1)(1,1)$ & $\left(0;0\right)$\\[1mm]
$\bar{Q} d$ &  $(1\oplus 8,2,\tfrac{1}{2})$ &$Y_D $ & $(3,1,\bar{3})(1,1)$ & $\left(0;0\right)$\\[1mm]
\hline 
$\bar{Q} Q^c$ &  $(6\oplus 3,1\oplus 3,\tfrac{1}{3})$ & $S_Q$ &$(6\oplus 3,1,1)(1,1)$ & $\left(\tfrac{2}{3};0\right)$  \\[1mm]
$\bar{u} d^c$ & $(6\oplus 3,1,\tfrac{1}{3})$ & $S_{UD}$ & $(1,3, 3)(1,1)$ & $\left(\tfrac{2}{3};0\right)$  \\[1mm]
$\bar{u} u^c$ &  $(6\oplus 3,1,\tfrac{4}{3})$ & $S_U$ &$(1,6\oplus 3,1)(1,1)$ & $\left(\tfrac{2}{3};0\right)$\\[1mm]
$\bar{d} d^c$ & $(6\oplus 3,1,-\tfrac{2}{3})$ & $S_D$ & $(1,1,6\oplus 3)(1,1)$ & $\left(\tfrac{2}{3};0\right)$\\[1mm]
$\bar{Q} \gamma^\mu u^c$ & $(6\oplus 3,2,\tfrac{5}{6})$ &  $\Delta_{QU}$ &$(3,3,1)(1,1)$ & $\left(\tfrac{2}{3};0\right)$\\[1mm]
$\bar{Q} \gamma^\mu d^c$ & $(6\oplus 3,2,-\tfrac{1}{6})$ & $\Delta_{QD}$ & $(3,1,3)(1,1)$ & $\left(\tfrac{2}{3};0\right)$\\[1mm]
\hline
\hline
$\bar{L}\gamma^\mu L$ &$(1,1\oplus 3,0)$ & $\Delta_L$ &  $(1,1,1)(1\oplus 8,1)$ & $\left(0;0\right)$\\[1mm]
$\bar{e} \gamma^\mu e$ &$(1,1,0)$ & $\Delta_E$ &  $(1,1,1)(1,1\oplus 8)$ & $\left(0;0\right)$\\[1mm]
$\bar{L} e$ &  $(1,2,\tfrac{1}{2})$ & $Y_E$ &$(1,1,1)(3,\bar{3})$ & $\left(0;0\right)$\\[1mm]
\hline 
$\bar{e}^c \gamma^\mu L$ &$(1,2,\tfrac{3}{2})$ & $\Delta_{EL}$ &  $(1,1,1)(\bar{3},\bar{3})$ & $\left(0;-2\right)$\\[1mm]
$\bar{L}^c L$ & $(1,1\oplus 3,1)$ &  $S_L$ &$(1,1,1)({\bar{6}}\oplus {\bar{3}},1)$ & $\left(0;-2\right)$\\[1mm]
$\bar{e}^c e$ & $(1,1,2)$ &  $S_E$ &$(1,1,1)(1,{\bar{6}}\oplus {\bar{3}})$ & $\left(0;-2\right)$\\[1mm]
\hline
\hline
$\bar{Q}\gamma^\mu L$ & $(3,1\oplus 3,\tfrac{2}{3})$ & $\Delta_{QL}$ &$(3,1,1)(\bar{3},1)$ & $\left(\tfrac{1}{3};-1\right)$ \\[1mm]
$\bar{u} \gamma^\mu e$ &  $(3,1,\tfrac{5}{3})$ & $\Delta_{UE}$ &$(1,3,1)(1,\bar{3})$ & $\left(\tfrac{1}{3};-1\right)$ \\ [1mm]
$\bar{d} \gamma^\mu e$ & $(3,1,\tfrac{2}{3})$ &  $\Delta_{DE}$ &$(1,1,3)(1,\bar{3})$ & $\left(\tfrac{1}{3};-1\right)$ \\[1mm]
$\bar{Q} e$ & $(3,2,\tfrac{7}{6})$ & $S_{QE}$ & $(3,1,1)(1,\bar{3})$ & $\left(\tfrac{1}{3};-1\right)$\\ [1mm]
$\bar{u} L$ & $(3,2,\tfrac{7}{6})$ & $S_{UL}$ & $(1,3,1)(\bar{3},1)$ & $\left(\tfrac{1}{3};-1\right)$ \\ [1mm]
$\bar{d} L$ & $(3,2,\tfrac{1}{6})$ & $S_{DL}$ & $(1,1,3)(\bar{3},1)$ & $\left(\tfrac{1}{3};-1\right)$\\ [1mm]
\hline 
$\bar{Q}^c \gamma^\mu e$ & $(\bar{3},2,\tfrac{5}{6})$ & $\Delta_{QE}$ & $(\bar{3},1,1)(1,\bar{3})$ & $\left(-\tfrac{1}{3};-1\right)$\\ [1mm]
$\bar{u}^c \gamma^\mu L$ &  $({\bar{3}},2,-\tfrac{1}{6})$ &$\Delta_{UL}$ & $(1,{\bar{3}},1)({\bar{3}},1)$ & $\left(-\tfrac{1}{3};-1\right)$ \\ [1mm]
$\bar{d}^c \gamma^\mu L$ & $(\bar{3},2,\tfrac{5}{6})$ & $\Delta_{DL}$ & $(1,1,\bar{3})(\bar{3},1)$ & $\left(-\tfrac{1}{3};-1\right)$ \\ [1mm]
$\bar{Q}^c L$ & $(\bar{3},1\oplus 3,\tfrac{1}{3})$ & $S_{QL}$ & $(\bar{3},1,1)(\bar{3},1)$ & $\left(-\tfrac{1}{3};-1\right)$\\[1mm]
$\bar{u}^c e$ & $(\bar{3},1,\tfrac{1}{3})$ &  $S_{UE}$ &$(1,\bar{3},1)(1,\bar{3})$ & $\left(-\tfrac{1}{3};-1\right)$\\[1mm]
$\bar{d}^c e$ &  $(\bar{3},1,\tfrac{4}{3})$ &$S_{DE}$ & $(1,1,\bar{3})(1,\bar{3})$ & $\left(-\tfrac{1}{3};-1\right)$\\[1mm]
\end{tabular}
}
\caption{Dirac bilinears and their associated flavour spurions, together with their quantum numbers under the SM gauge symmetries and the flavour group ${\cal{G}}_f$. Baryon and lepton numbers of the spurions are also listed.}
\label{tab:1}
\end{center}
\end{table}

In principle, each of the spurions of \Table{tab:1} can contribute to the 
non-diagonal flavour structure of different SMEFT operators. 
In particular, if we identify the flavour structure as originating from the tree-level exchange of heavy particles,
the flavour coefficients of the 4-fermion operators in SMEFT would factorise as a product of the corresponding
spurions, as explained already in the Introduction.

Given the indications that NP effects might already be present in semileptonic $B$ decays, one can concentrate on the subset of operators with two quarks and two leptons. There are 10 such operators, which are listed in \Table{tab:5}. Each of the operators comes with a dimensional suppression and possesses flavour structure. For instance, the first operator in \Table{tab:5} appears in the effective Lagrangian as
\begin{align}
\frac{1}{\Lambda^2}[{\mathcal{C}}_{lq}]^{ij\alpha\beta}({\bar{Q}}_i\gamma_{\mu}Q_j)({\bar{L}}_{\alpha}\gamma^{\mu}L_{\beta}) \,,
\end{align} 
and corresponds to the example used in the Introduction.

\Table{tab:5} allows one to identify the relevant spurions that would be needed to generate a non-trivial
flavour coefficient for a particular 4-fermion  SMEFT operator. 
In the next Sections we will concentrate on a scenario where only $\Delta_{QL}$ and $\Delta_{DE}$ are 
present, which singles out the first, second, fifth and eighth line in \Table{tab:5}.

\begin{table}[hbt!]
\begin{center}
\centering
\renewcommand{\arraystretch}{1.2} 
\begin{tabular}{l| l}
\qquad\qquad Operator & \qquad\qquad\qquad\qquad\qquad  Coefficient \\[1mm]
\hline\\[-4mm]
$({\bar{Q}}^i\gamma_{\mu}Q^j)({\bar{L}}^{\alpha}\gamma^{\mu}L^{\beta})$ & $[{\mathcal{C}}_{lq}^{(1)}]^{ij\alpha\beta}=a_0[\Delta_Q]^{ij}[\Delta_L]^{\alpha\beta}+b_0[\Delta_{QL}]^{i\alpha}[\Delta_{QL}^{\dagger}]^{j\beta}+c_0[S_{QL}]^{i\alpha}[S_{QL}^{\dagger}]^{j\beta}$ \\[1mm]
$({\bar{Q}}^i\gamma_{\mu}\tau^a Q^j)({\bar{L}}^{\alpha}\gamma^{\mu}\tau^aL^{\beta})$ & $[{\mathcal{C}}_{lq}^{(3)}]^{ij\alpha\beta}=a_1[\Delta_Q]^{ij}[\Delta_L]^{\alpha\beta}+b_1[\Delta_{QL}]^{i\alpha}[\Delta_{QL}^{\dagger}]^{j\beta}+c_1[S_{QL}]^{i\alpha}[S_{QL}^{\dagger}]^{j\beta}$\\[1mm]
$({\bar{Q}}^i\gamma_{\mu}Q^j)({\bar{e}}^{\alpha}\gamma^{\mu}e^{\beta})$ & $[{\mathcal{C}}_{eq}]^{ij\alpha\beta}=a_2[\Delta_Q]^{ij}[\Delta_E]^{\alpha\beta}+b_2[\Delta_{QE}]^{i\alpha}[\Delta_{QE}^{\dagger}]^{j\beta}+c_2[S_{QE}]^{i\alpha}[S_{QE}^{\dagger}]^{j\beta}$\\[1mm]
$({\bar{d}}^i\gamma_{\mu}d^j)({\bar{L}}^{\alpha}\gamma^{\mu}L^{\beta})$ & $[{\mathcal{C}}_{ld}]^{ij\alpha\beta}=a_3[\Delta_D]^{ij}[\Delta_L]^{\alpha\beta}+b_3[\Delta_{DL}]^{i\alpha}[\Delta_{DL}^{\dagger}]^{j\beta}+c_3[S_{DL}]^{i\alpha}[S_{DL}^{\dagger}]^{j\beta}$\\[1mm]
$({\bar{d}}^i\gamma_{\mu}d^j)({\bar{e}}^{\alpha}\gamma^{\mu}e^{\beta})$ & $[{\mathcal{C}}_{ed}]^{ij\alpha\beta}=a_4[\Delta_D]^{ij}[\Delta_E]^{\alpha\beta}+b_4[\Delta_{DE}]^{i\alpha}[\Delta_{DE}^{\dagger}]^{j\beta}+c_4[S_{DE}]_{i\alpha}[S_{DE}^{\dagger}]^{j\beta}$\\[1mm]
$({\bar{u}}^i\gamma_{\mu}u^j)({\bar{L}}^{\alpha}\gamma^{\mu}L^{\beta})$ & $[{\mathcal{C}}_{lu}]^{ij\alpha\beta}=a_5[\Delta_U]^{ij}[\Delta_L]^{\alpha\beta}+b_5[\Delta_{UL}]^{i\alpha}[\Delta_{UL}^{\dagger}]^{j\beta}+c_5[S_{UL}]^{i\alpha}[S_{UL}^{\dagger}]^{j\beta}$\\[1mm]
$({\bar{u}}^i\gamma_{\mu}u^j)({\bar{e}}^{\alpha}\gamma^{\mu}e^{\beta})$ & $[{\mathcal{C}}_{eu}]^{ij\alpha\beta}=a_6[\Delta_U]^{ij}[\Delta_E]^{\alpha\beta}+b_6[\Delta_{UE}]^{i\alpha}[\Delta_{UE}^{\dagger}]^{j\beta}+c_6[S_{UE}]^{i\alpha}[S_{UE}^{\dagger}]^{j\beta}$\\[1mm]
$({\bar{d}}^iQ^j)({\bar{L}}^{\alpha}e^{\beta})$ & $[{\mathcal{C}}_{ledq}]^{ij\alpha\beta}=a_7[Y_D^{\dagger}]^{ij}[Y_E]^{\alpha\beta}+b_7[\Delta_{DE}]^{i\alpha}[\Delta_{QL}^{\dagger}]^{j\beta}+c_7[S_{DE}]^{i\alpha}[S_{QL}^{\dagger}]^{j\beta}$\\[1mm]
$\epsilon_{ab}({\bar{Q}}^i_au^j)({\bar{L}}^{\alpha}_be^{\beta})$ & $[{\mathcal{C}}_{lequ}^{(1)}]^{ij\alpha\beta}=a_8[Y_U]_{ij}[Y_E]^{\alpha\beta}+b_8[\Delta_{QE}]^{i\alpha}[\Delta_{UL}^{\dagger}]^{j\beta}+c_8[S_{QE}]^{i\alpha}[S_{UL}^{\dagger}]^{j\beta}$\\[1mm]
$\epsilon_{ab}({\bar{Q}}^i_a\sigma_{\mu\nu}u^j)({\bar{L}}^{\alpha}_b\sigma^{\mu\nu}e^{\beta})$ & $[{\mathcal{C}}_{lequ}^{(3)}]^{ij\alpha\beta}=a_{9}[Y_U]_{ij}[Y_E]^{\alpha\beta}+b_{9}[\Delta_{QE}]^{i\alpha}[\Delta_{UL}^{\dagger}]^{j\beta}+c_{9}[S_{QE}]^{i\alpha}[S_{UL}^{\dagger}]^{j\beta}$
\end{tabular}
\caption{Decomposition of the 4-fermion SMEFT flavour coefficients in terms of spurions. Here we did not list 
higher orders in the spurion expansion that could be induced by radiative corrections.}
\label{tab:5}
\end{center}
\end{table}


\subsection{Consistency conditions and Froggatt-Nielsen charges}

\label{sec:FN}

As already mentioned above, in an EFT approach with non-trivial flavour structures one 
has to satisfy certain consistency conditions. They ensure that hierarchical patterns 
which one assumes for the flavour structure associated to one effective operator are not 
spoiled by the combinations of flavour structures appearing in any of the other operators
\cite{Feldmann:2006jk}.
In particular, this is a necessary requirement to ensure that the flavour hierarchies are 
stable under renormalisation-group evolution.  For instance, in the SM the 
following inequality holds,
\begin{align}
  |(Y_U)^{ij}| \geq |(Y_D Y_D^\dagger Y_U)^{ij}| \,,
\end{align}
where the matrices on both sides transform in the same way under the flavour symmetry ${\cal G}_f$.

In the SM, all consistency relations of the above type 
hold trivially, because all eigenvalues of $Y_U$ and $Y_D$ are smaller (or equal) than one,
and the CKM angles satisfy relations of the type $\theta_{13} \theta_{12} \leq \theta_{23}$. 
They also hold in the MFV approach, since no additional flavour structures apart from the SM Yukawa matrices appear. However, as soon as one includes new flavour structures, the consistency relations become a non-trivial 
theoretical requirement (see the discussion in \cite{Feldmann:2006jk}).

As mentioned in the Introduction, an efficient way to fulfil all the consistency conditions is to use Froggatt-Nielsen (FN) charges \cite{Froggatt:1978nt} to define a power-counting scheme for arbitrary flavour structures. We will denote as $b_Q^i, b_D^i, b_U^i$ and $b_L^\alpha, b_E^\alpha$ the FN charges for the fermions in a flavour basis defined by the $U(1)$ symmetry of the FN construction (FN basis). The entries of the Yukawa matrices then scale with the small parameter $\lambda \sim 0.2 $ as 
\begin{align}
(Y_U)_{ij}\sim& \lambda^{\vert b_Q^i-b_U^j\vert}\,, \qquad  
(Y_D)_{ij}\sim \lambda^{\vert b_Q^i-b_U^j\vert}\,,\qquad  
(Y_E)_{\alpha\beta}\sim \lambda^{\vert b_L^\alpha-b_E^\beta \vert}\,.
\end{align}
In this case, the above example for a consistency condition would simply  
translate into 
\begin{equation}
  \lambda^{|b_Q^i - b_U^j|} \geq 
  \lambda^{|b_Q^i - b_D^k| + |b_D^k-b_Q^l| + |b_Q^l - b_U^j|}\,,
\end{equation}
which is true because of triangle inequalities. Analogous relations would then also hold for 
products of arbitrary flavour spurions with the flavour structure fixed by universal FN charges.
For instance, in the $U_1$ vector-leptoquark scenario, to be further discussed below, 
one would consider the additional
spurions $\Delta_{QL}$ and $\Delta_{DE}$. For these the following inequalities 
hold 
 \begin{align}
  |Y_E^{\alpha\beta}| & \lesssim \left| \left(\Delta_{QL}^\dagger Y_D \Delta_{DE}\right)^{\alpha\beta} \right| \,, 
  \quad & 
  |Y_D^{ij}|  & \lesssim \left| \left(\Delta_{QL} Y_E \Delta_{DE}^\dagger\right)^{ij} \right| \,,
  \cr 
  |\Delta_{DE}^{i\alpha}|  & \lesssim \left| \left(Y_D^\dagger \Delta_{QL} Y_E \right)^{i\alpha} \right| \,,
  \quad &
  |\Delta_{QL}^{i\alpha}|  & \lesssim \left| \left(Y_D \Delta_{DE} Y_E^\dagger\right)^{i\alpha} \right| \,.
 \end{align}
Notice that these inequalities are to be understood in the FN basis.

Since the FN power counting has to reproduce the SM flavour hierarchies, some of the FN charges are fixed from the known fermion masses and CKM mixing angles. Concerning the latter, the FN power counting yields
\begin{align}
(V_{\rm CKM})_{ij}=(V_{U_L}^{\dagger}V_{D_L})_{ij} 
\sim \lambda^{\vert b_Q^i-b_Q^j\vert} \,, 
\end{align}
where $V_X$ denote the rotation matrices from the flavour to the mass eigenbasis for a given fermion species. Comparing this with the generally accepted Wolfenstein power-counting for the CKM matrix,
\begin{equation}\label{CKM}
V_{\rm CKM}\sim \left(
\begin{array}{ccc}
1 & \lambda & \lambda^3\\
\lambda & 1 & \lambda^2\\
\lambda^3 & \lambda^2 & 1
\end{array}
\right)\,,
\end{equation}
determines the FN charges $b_Q^i$ up to a common offset $d$ and an absolute sign. There are two families of general solutions:
\begin{align}
b_Q^1=3+d\,, \qquad b_Q^2=2+d\,, \qquad b_Q^3=d\,
\end{align}
and 
\begin{align}
b_Q^1=3+d\,, \qquad b_Q^2=4+d\,, \qquad b_Q^3=6+d\,.
\end{align}
Since in this work we only consider flavour structures that are associated with Dirac currents $\bar \psi  \Gamma \psi$, the 
offset is irrelevant for our discussion. For definiteness we will set 
\begin{align}
 b_Q^1 \equiv 3 \,, \qquad b_Q^2 \equiv 2 \,, \qquad b_Q^3 \equiv 0 \,.
\end{align}
Concerning the eigenvalues of the up-quark Yukawa matrix, we have 
\begin{align}
 y_u & \sim \lambda^{|b_Q^1-b_U^1|} \approx \lambda^8 \,, 
 \cr 
 y_c& \sim \lambda^{|b_Q^2-b_U^2|} \approx \lambda^4 \,, 
 \cr 
 y_t & \sim \lambda^{|b_Q^3-b_U^3|} \approx \lambda^0 \,, 
\end{align}
Here and in the following, the integers on the right-hand side have to be understood as an estimate: deviations by one unit in the exponent are not excluded. The above expressions fix the charges $b_U^i$ up to a $2^2$-fold ambiguity:
\begin{align}
b_U^1 \simeq-5 (+11)  \,, 
\qquad 
b_U^2 \simeq-2 (+6) \,, 
\qquad 
b_U^3 \equiv 0  \,.
\end{align}
The same analysis for the down-type quarks yields 
\begin{align}
 y_d & \sim \lambda^{|b_Q^1-b_D^1|} \approx \lambda^7 \,, 
 \cr 
 y_s& \sim \lambda^{|b_Q^2-b_D^2|} \approx \lambda^5 \,, 
 \cr 
 y_b & \sim \lambda^{|b_Q^3-b_D^3|} \approx \lambda^3 \,, 
\end{align}
and, consequently, 
\begin{align}
b_D^1 \simeq-4 (+10)  \,, 
\qquad 
b_D^2 \simeq-3 (+7) \,, 
\qquad 
b_D^3 \equiv -3(+3)  \,.
\end{align}
Notice that the FN charges for the right-handed quark singlets also determine the scaling 
of the corresponding rotation matrices (which are not observable in the SM),
\begin{align}
(V_{U})_{ij} =
(V_{U_R})_{ij}
\sim \lambda^{\vert b_U^i-b_U^j\vert}  \,, \qquad 
(V_D)^{ij} =
(V_{D_R})_{ij}
\sim \lambda^{\vert b_D^i-b_D^j\vert} \,.
\end{align}
In this way, it is guaranteed that the power-counting for the NP flavour spurions does not change when rotating from the flavour to the mass eigenbasis of the SM fermions. This is another advantageous feature of the FN power counting.

In the lepton sector, the experimental knowledge of the FN charges is sparse. The masses of the charged leptons can be used to estimate only three charge \emph{differences}, namely
\begin{align}
 y_e & \sim \lambda^{|b_L^1-b_E^1|} \approx \lambda^9 \,, 
 \cr 
 y_\mu& \sim \lambda^{|b_L^2-b_E^2|} \approx \lambda^5 \,, 
 \cr 
 y_\tau & \sim \lambda^{|b_L^3-b_E^3|} \approx \lambda^3 \,.
\end{align}
Therefore, in order to comply with the SM flavour structure there is a lot of freedom left on the choice of FN charges. The different allowed combinations of charges however lead to very different predictions for NP flavour structures. This of course depends on the version of extended MFV that one selects. In the next Section we will discuss a minimal extension of MFV able to accommodate the $B$ anomalies and then work out the associated constraints on the FN charges.


\section{Extended MFV from $U_1$ Vector Leptoquark}
\label{sec:U1_model}

As already mentioned in the previous Sections, a promising scenario to address the $B$ anomalies without generating tensions with Electroweak Precision Tests (EWPT) or high-$p_T$ observables \cite{Buttazzo:2017ixm} is to postulate the existence of a vector leptoquark $U_1$. This scenario has to be understood as a simplified model, which requires a UV completion. In the literature there exist a number of different proposals trying to achieve such a task, see e.g. \cite{Bordone:2018nbg,Bordone:2017bld,Cornella:2019hct,Blanke:2018sro,DiLuzio:2017vat,DiLuzio:2018zxy}. Our approach in this paper will be much more modest: we are not interested in the  
\emph{dynamics} of the leptoquark scenario or its possible UV completion, but rather 
concentrate on the imprint that such dynamics could have on the \emph{flavour structures} observed at low energies.
In the context of SMEFT we thus use the leptoquark model as a criterium to select the relevant flavour spurions 
discussed in \sec{sec:2}. Whether this approach results in a viable candidate to accommodate flavour
observables can then be studied in a model-independent way.

For the sake of this work, we will adopt further simplifying assumptions and only concentrate 
on SMEFT operators that catch the leading effects of leptoquark couplings to SM fermions.
In most cases, this amounts to assuming tree-level relations for flavour coefficients that arise from 
leptoquark exchange, with the exception of LFU ratios for $W$ couplings, where we take into account
one-loop results (see below). 

\subsection{The simplified $U_1$ scenario}
\label{sec:simp}
The flavour-specific interactions between $U_1$ 
and SM fermions are described 
by the introduction of two spurions:
 \begin{equation}
\label{eq:simpl_U1_lagr}
\mathcal{L} = \Delta_{QL} ^{i\alpha} \left(  
\bar{Q}^i\gamma_\mu L^\alpha \right)  U_1^\mu +
\Delta_{DE}^{i\alpha} \left( 
\bar{d}^i \gamma_\mu e^\alpha \right) U_1^\mu+ \text{h.c.} 
\end{equation}
With the FN power counting, the spurions $ \Delta_{QL}^{i\alpha}$ and $ \Delta_{DE}^{i\alpha}$
can be parameterized as 
\begin{align}
\label{eq:spurionstoWC}
\Delta_{QL}^{i\alpha}& =  c_{QL}^{i\alpha}\lambda^{|b_Q^i-b_L^\alpha|}  \,, \\
\Delta_{DE}^{i\alpha} &=   c_{DE}^{i\alpha} \lambda^{|b_D^i-b_E^\alpha|} \,.
\end{align}
Here $c_{QL(DE)}^{i\alpha}$ are flavour-dependent coefficients of $\mathcal{O}(1)$.
In general, these coefficients could be complex and carry additional CP-violating phases.
Since we are not considering CP-violating observables in this work, we make a further simplifying
assumption and take all the coefficients real in a basis where the Yukawa matrices for 
down quarks and charged leptons are diagonal and real. In this way the CKM matrix remains the only 
source of CP violation in the flavour sector.

Once we integrate out the $U_1$ leptoquark, we get the following contributions to the 
relative 4-fermion operators in the effective dim-6 Lagrangian:
\begin{equation}
\label{eq:dim6U1}
\begin{aligned}
\mathcal{L}_\text{eff}=\mathcal{L}_\text{SM}-\frac{1}{\Lambda^2}\bigg\{ &[\mathcal{C}^{(3)}_{lq}]^{ij\alpha\beta} (\bar{Q}^i\gamma^\mu \sigma^a Q^j)(\bar{L}^\alpha\gamma_\mu \sigma^a L^\beta)+[\mathcal{C}^{(1)}_{lq}]^{ij\alpha\beta} (\bar{Q}^i\gamma^\mu Q^j)(\bar{L}^\alpha\gamma_\mu L^\beta) \\
+& [\mathcal{C}_{ed}]^{ij\alpha\beta}(\bar{d}^i\gamma^\mu d^j)(\bar{e}^\alpha\gamma_\mu e^\beta)+[\mathcal{C}_{ledq}]^{ij\alpha\beta}(\bar{Q}^i d^j)(\bar{e}^\alpha L^\beta) + \text{h.c.} \bigg\} \,,
\end{aligned}
\end{equation}
where $\Lambda$ is an effective scale associated with the leptoquark mass.
In the broken phase, the Lagrangian in \eq{eq:dim6U1} acquires the form shown in \eq{eq:dim6U1broken}. 
The basis chosen for the $SU(2)_L$ quark and lepton doublets is the down-quark basis described in \eq{eq:SU_basis}.

The tree-level matching relations
between the SMEFT Wilson coefficients and the spurions from the leptoquark couplings
are easy to find and read 
\begin{align}
[\mathcal{C}^{(1)}_{lq}]^{ij\alpha\beta} = [\mathcal{C}^{(3)}_{lq}] ^{ij\alpha\beta} =&\,+ \Delta_{QL}^{i\alpha}\,\Delta_{QL}^{*j\beta} \,,\label{eq:matching_C1C3} \\
[\mathcal{C}_{leqd}]^{ij\alpha\beta} =&\,   -2\,\Delta_{QL}^{i\alpha} \, \Delta_{DE}^{* j\beta}  \,,    \\
[\mathcal{C}_{ed}]^{ij\alpha\beta} =&\, +\Delta_{DE}^{i\alpha} \, \Delta_{DE}^{* j\beta} \,.
\end{align}
Notice that the relation in \eq{eq:matching_C1C3} is a tree-level result and, once a UV completion of the simplified model is specified, gets modified by higher-order radiative corrections. Since (\ref{eq:matching_C1C3}) is not a fundamental relation, from a bottom-up approach the weak singlet and triplet coefficients, $\mathcal{C}^{(1)}_{lq}$ and $\mathcal{C}^{(3)}_{lq}$, should in general be treated as independent coefficients.

In any case, the power counting for the individual entries is dictated by FN charges, i.e.\
our approach leads to definite predictions for the order of magnitude of NP effects in all possible 
flavour transitions which could be mediated by the $U_1$ leptoquark.  
It then remains the task to assess the phenomenological consistency of
this approach. To this end we will perform a rather exhaustive fit including various flavour and precision observables.

\subsection{Relevant low-energy observables}
\label{sec:relevant}
In the following, we list the expressions for the whole set of observables we will employ in the fit of \sec{sec:fit} in terms of the effective operators at the hadronic scale, with the FN power counting made explicit. 
The matching to SMEFT Wilson coefficients can be found in \app{app:2}.

\subsubsection{FCNC-mediated processes}
\label{sec:FCNCs}
The generic Wilson coefficients relevant for processes involving FCNCs are 
\begin{align}
\mathcal{C}_{9}^{ij\alpha\beta} =& -\,\mathcal{C}_{10}^{ij\alpha\beta} =\frac{2\,v^2}{\Lambda^2}\frac{\pi}{\alpha_\text{EM} \left|V_{tb}V^*_{ts}\right|} \, c_{QL}^{i\alpha} \lambda^{|b_Q^i-b_L^{\alpha}|}\, c_{QL}^{j\beta} \lambda^{|b_Q^j-b_L^{\beta}|}\,,\label{eq:genV1} \\
\mathcal{C}_{9}^{\prime\, ij\alpha\beta} =& + \,\mathcal{C}_{10}^{\prime\, ij\alpha\beta} =\frac{\,v^2}{\Lambda^2}\frac{\pi}{\alpha_\text{EM} \left|V_{tb}V^*_{ts}\right|}\, c_{DE}^{i\alpha} \lambda^{|b_D^i-b_E^{\alpha}|} \, c_{DE}^{j\beta} \lambda^{|b_D^j-b_E^{\beta}|}\,,\label{eq:genV2}\\
\mathcal{C}_{S}^{ij\alpha\beta} =& - \,\mathcal{C}_{P}^{ij\alpha\beta} =\frac{2\,v^2}{\Lambda^2}\frac{\pi}{\alpha_\text{EM} \left|V_{tb}V^*_{ts}\right|} \, c_{QL}^{i\alpha} \lambda^{|b_Q^i-b_L^{\alpha}|} \, c_{DE}^{j\beta} \lambda^{|b_D^j-b_E^{\beta}|}\,, \label{eq:genV3}\\
\mathcal{C}_{S}^{\prime\, ij\alpha\beta} =& + \,\mathcal{C}_{P}^{\prime\, ij\alpha\beta} =\frac{2\,v^2}{\Lambda^2}\frac{\pi}{\alpha_\text{EM} \left|V_{tb}V^*_{ts}\right|} \, 
c_{DE}^{i\alpha} \lambda^{|b_D^i-b_E^{\alpha}|} \, c_{QL}^{j\beta} \lambda^{|b_Q^j-b_L^{\beta}|}\,. \label{eq:genV4}
\end{align}
The relations ${\mathcal{C}}_S=-{\mathcal{C}}_P$ and ${\mathcal{C}}_S^{\prime}={\mathcal{C}}_P^{\prime}$ are a consequence of having a SM Higgs, i.e. a scalar weak doublet \cite{Alonso:2014csa,Cata:2015lta}, while the relations between ${\cal{C}}_9^{(\prime)}$ and ${\cal{C}}_{10}^{(\prime)}$ are specific of the $U_1$ model. In the following we will explicitly spell out the contributions to the different processes considered, including Lepton Flavour Violating (LFV) decays.
\begin{itemize}
\item[(a)]  \boldmath$b\to s\mu^+\mu^-\unboldmath$ \textbf{modes}\unboldmath.
The contributions to $b\to s\mu^+\mu^-$ can be derived from \eq{eq:observables_bsll}
and involve the flavour entries
\begin{align}
\mathcal{C}_{9}^{2322} = - \,\mathcal{C}_{10}^{2322} \,,\qquad \qquad
\mathcal{C}_{9}^{\prime\, 2322} = + \,\mathcal{C}_{10}^{\prime\, 2322} \,,
\end{align}
for the vector and axial-vector operators, and 
\begin{align}
\mathcal{C}_{S}^{2322} = - \,\mathcal{C}_{P}^{2322}\,,\qquad \qquad
\mathcal{C}_{S}^{\prime\, 2322} = + \,\mathcal{C}_{P}^{\prime\, 2322}\,,
\label{eq:bsmumu_FN}
\end{align}
for the scalar and pseudoscalar operators.
Results of global fits require $\mathcal{C}_{9}^{2322}<0$ and $\mathcal{C}_{9}^{\prime 2322}>0$
\cite{Alguero:2019ptt,Aebischer:2019mlg,Ciuchini:2019usw,Alok:2019ufo}, which translate into the conditions $c_{QL}^{32} c_{QL}^{22}<0$ and $c_{DE}^{22}c_{DE}^{32}>0$, respectively. Since the values for $\mathcal{C}_{9}^{\prime 2322}$ are small, the latter condition will not be relevant for our analysis.   
We will comment more on the implications of these conditions in \sec{sec:fit}.
\item[(b)] \boldmath$\bar B_d\to \ell_i^+\ell_j^-$\unboldmath\, {\bf{decay modes}}. The contributions to these decays are:
\begin{align}
\bar B_d\to e^+\mu^-:  \qquad \qquad {\mathcal{C}}^{1312}\,,\\
\bar B_d\to e^-\mu^+:  \qquad \qquad {\mathcal{C}}^{1321}\,,\\
\bar B_d\to e^+\tau^-:  \qquad \qquad {\mathcal{C}}^{1313}\,,\\
\bar B_d\to\mu^+\tau^-:  \qquad \qquad {\mathcal{C}}^{1323}\,,
\end{align} 
where ${\cal{C}}$ is a short-hand notation to denote collectively the 8 different Wilson coefficients that contribute to each decay mode. This notation is also used in the next two points.

At present no experimental bounds exist for the decay modes $\bar B_d\to\mu^-\tau^+$ and $\bar B_d\to e^-\tau^+$.
\item[(c)] \boldmath$\bar B_s\to \ell_i^+\ell_j^-$\unboldmath\, {\bf{decay modes}}. The relevant coefficients in this case are:
\begin{align}
\bar B_s\to e^+\mu^-:  \qquad \qquad {\mathcal{C}}^{2312}\,,\\
\bar B_s\to e^-\mu^+:  \qquad \qquad {\mathcal{C}}^{2321}\,,\\
\bar B_s\to\mu^+\tau^-:  \qquad \qquad {\mathcal{C}}^{2323}\,,\\
\bar B_s\to\mu^-\tau^+:  \qquad \qquad {\mathcal{C}}^{2332}\,,\\
\bar B_s\to\tau^+\tau^-:  \qquad \qquad {\mathcal{C}}^{2333}\,.
\end{align} 
\item[(d)]\boldmath$K_L\to \ell_i^+\ell_j^-$\unboldmath\, {\bf{decay modes}}. For these modes one finds:
\begin{align}
K_L\to e^+\mu^-: \qquad \qquad {\mathcal{C}}^{2112}\,,\\
K_L\to e^-\mu^+: \qquad \qquad {\mathcal{C}}^{2121}\,.
\end{align}
\end{itemize}


\subsubsection{$R_{D^{(*)}}$}

The contributions to $R_{D^{(*)}}$ are sensitive to  the operators 
$\mathcal{O}_{lq}^{(3)}$ and $\mathcal{O}_{ledq}$. 
From \eqs{eq:btoc_coeffs1}{eq:observable: RDst} we read off
\begin{align}
\label{eq:RD_complete}
R_{D^{(*)}}= R_{D^{(*)}}^{\text{SM}}&\left[1+ \frac{2\,v^2}{\Lambda^2}c_{QL}^{33} \lambda^{|b_Q^3-b_L^3|}\Re\left(c_{QL}^{33}\lambda^{|b_Q^3-b_L^3|}+c_{QL}^{23}\frac{V_{cs}}{V_{cb}} \lambda^{|b_Q^2-b_L^3|} +c_{QL}^{13}\frac{V_{cd}}{V_{cb}}\lambda^{|b_Q^1-b_L^3|}\right)\right. \notag\\
&-\frac{2\,v^2}{\Lambda^2} f_{D^{(*)}}^S (\tau)\, c_{DE}^{33} \lambda^{|b_D^3-e_L^3|}\Re\left(c_{QL}^{33}\lambda^{|b_Q^3-b_L^3|}+c_{QL}^{23}\frac{V_{cs}}{V_{cb}} \lambda^{|b_Q^2-b_L^3|} +c_{QL}^{13}\frac{V_{cd}}{V_{cb}}\lambda^{|b_Q^1-b_L^3|}\right) \notag\\
&\left.- \frac{2\,v^2}{\Lambda^2} c_{QL}^{32}\lambda^{|b_Q^3-b_L^2|}\Re\left(c_{QL}^{32}\lambda^{|b_Q^3-b_L^2|}+c_{QL}^{22}\frac{V_{cs}}{V_{cb}} \lambda^{|b_Q^2-b_L^2|} +c_{QL}^{12}\frac{V_{cd}}{V_{cb}}\lambda^{|b_Q^1-b_L^2|}\right)\right] \,,
\end{align}
where $f_{D^*}^S(\tau)=0.12$ and $f_{D}^S(\tau)=1.5$ are the integrated form factors given in \cite{Fajfer:2012vx,Lattice:2015rga,Na:2015kha}. Contributions proportional to  $f_{D^{(*)}}^S(\mu)$ are numerically negligible. According to our criterium of keeping only the (dominant) interference of NP operators with the SM ones, in the expression above we have neglected LFV contributions.

\subsubsection{Universality of $|V_{cb}|$}

NP contributions give in general different corrections to the CKM element $|V_{cb}|$, depending on the decay channel probed. For $b\to c \ell \nu$, and denoting by $\tilde{V}^{\ell}_{cb}$ the associated correction, one finds
\be
\begin{aligned}
\frac{\tilde{V}^e_{cb}}{\tilde{V}^\mu_{cb}}=  1&+  \frac{v^2}{\Lambda^2}\, c_{QL}^{31}\lambda^{ |b_Q^3-b_L^1|}\left( c_{QL}^{31}\lambda^{|b_Q^3-b_L^1|}+c_{QL}^{21}\frac{V_{cs}}{V_{cb}}\lambda^{|b_Q^2-b_L^1|}+c_{QL}^{11}\frac{V_{cd}}{V_{cb}}\lambda^{|b_Q^1-b_L^1|}\right) \\
&-\frac{v^2}{\Lambda^2}\, c_{QL}^{32}\lambda^{|b_Q^3-b_L^2|}\left(c_{QL}^{32}\lambda^{|b_Q^3-b_L^2|}+c_{QL}^{22}\frac{V_{cs}}{V_{cb}}\lambda^{|b_Q^2-b_L^2|}+c_{QL}^{12}\frac{V_{cd}}{V_{cb}}\lambda^{|b_Q^1-b_L^2|} \right)\,.
\end{aligned}
\ee
LFV contributions do not interfere with the SM and have been neglected. Furthermore, contributions from scalar operators have a $m_{\ell}$ chiral suppression and can be safely dismissed.

\subsubsection{Radiative decays}
\label{sec:3.4}
Constraints on both $Z$ and $W$ couplings are also induced at the one-loop level. For the $Z$ coupling, the main correction affects its coupling with neutrinos via a top loop. Using \eq{eq:rad_Znu} 
we find
\begin{equation}
\left(\delta g^Z_{\nu_L}\right)_{\alpha\beta} = \frac{1}{16\pi^2}\frac{v^2}{\Lambda^2} 3 y_t^2 \left( 2\, c_{QL}^{3\alpha} c_{QL}^{3\beta}\lambda^{b_Q^3-b_L^\alpha} \lambda^{b_Q^3-b_L^\beta}\right)L_t \,.
\label{eq:rad_Znu_FN}
\end{equation}
Corrections to $Z\to \tau^+\tau^-$ vanish if $\mathcal{C}_{lq}^{(1)}=\mathcal{C}_{lq}^{(3)}$, as we are assuming.

For the $W$ vertices, the corrections to the leptonic decays read
\be
\begin{aligned}
\left(\frac{g_\tau}{g_\mu}\right)_\text{lept}=& \,1-\frac{v^2}{\Lambda^2} \frac{3 y_t^2}{8\pi^2}(c_{QL}^{33})^2\lambda^{2 |b_Q^3-b_L^3|}L_t+\frac{v^2}{\Lambda^2} \frac{3 y_t^2}{8\pi^2}(c_{QL}^{32})^2\lambda^{2 |b_Q^3-b_L^2|}L_t \,,  \\
\left(\frac{g_\tau}{g_e}\right)_\text{lept}=&\, 1- \frac{v^2}{\Lambda^2} \frac{3 y_t^2}{8\pi^2}(c_{QL}^{33})^2\lambda^{2 |b_Q^3-b_L^3|}L_t+\frac{v^2}{\Lambda^2} \frac{3 y_t^2}{8\pi^2}(c_{QL}^{31})^2\lambda^{2 |b_Q^3-b_L^1|}L_t \,,  \\
\left(\frac{g_\mu}{g_e}\right)_\text{lept}=& \, 1-  \frac{v^2}{\Lambda^2} \frac{3 y_t^2}{8\pi^2}(c_{QL}^{32})^2\lambda^{2 |b_Q^3-b_L^2|}L_t+\frac{v^2}{\Lambda^2} \frac{3 y_t^2}{8\pi^2}(c_{QL}^{31})^2\lambda^{2 |b_Q^3-b_L^1|}L_t \,,
\end{aligned}
\ee
while for the hadronic modes one finds
\begin{align}
\left(\frac{g_\tau}{g_\mu}\right)_\pi =  \, 1+&\, \frac{2v^2}{\Lambda^2} \,c_{QL}^{13} \lambda^{ |b_Q^1-b_L^3|}\left( c_{QL}^{13} \lambda^ {|b_Q^1-b_L^3|}+c_{QL}^{23}\frac{V_{us}}{V_{ud}} \lambda^{|b_Q^2-b_L^3|}+c_{QL}^{33}\frac{V_{ub}}{V_{ud}}\lambda^{|b_Q^3-b_L^3|}\right) \notag\\
 -&\frac{2v^2}{\Lambda^2} \, c_{QL}^{12} \lambda^{|b_Q^1-b_L^2|}\left(c_{QL}^{12} \lambda^{|b_Q^1-b_L^2|}+c_{QL}^{22}\frac{V_{us}}{V_{ud}} \lambda^{ |b_Q^2-b_L^2|}+c_{QL}^{32}\frac{V_{ub}}{V_{ud}} \lambda^{ |b_Q^3-b_L^2|} \right) \\
 - &\frac{2v^2}{\Lambda^2}  \frac{m_\pi^2}{m_\tau (m_u+m_{d})}c_{DE}^{13} \lambda^{ |b_D^1-b_E^3|}\left( c_{QL}^{13} \lambda^ {|b_Q^1-b_L^3|}+c_{QL}^{23}\frac{V_{us}}{V_{ud}} \lambda^{|b_Q^2-b_L^3|}+c_{QL}^{33}\frac{V_{ub}}{V_{ud}}\lambda^{|b_Q^3-b_L^3|}\right)\notag \\
  +&\frac{2v^2}{\Lambda^2}  \frac{m_\pi^2}{m_\mu (m_u+m_{d})}c_{DE}^{12} \lambda^{ |b_D^1-b_E^2|}\left( c_{QL}^{12} \lambda^ {|b_Q^1-b_L^2|}+c_{QL}^{22}\frac{V_{us}}{V_{ud}} \lambda^{|b_Q^2-b_L^2|}+c_{QL}^{32}\frac{V_{ub}}{V_{ud}}\lambda^{|b_Q^3-b_L^2|}\right)\,,\notag \\
\left(\frac{g_\tau}{g_\mu}\right)_K =  \, 1+& \, \frac{2v^2}{\Lambda^2} \,c_{QL}^{23}\lambda^{ |b_Q^2-b_L^3|}  \left(c_{QL}^{23}\lambda^{ |b_Q^2-b_L^3|}+c_{QL}^{13} \frac{V_{ud}}{V_{us}}\lambda^{|b_Q^1-b_L^3|}+c_{QL}^{33}\frac{V_{ub}}{V_{us}}\lambda^{|b_Q^3-b_L^3|}\right) \notag \\
-& \frac{2v^2}{\Lambda^2} \,c_{QL}^{22}\lambda^{|b_Q^2-b_L^2|} \left(c_{QL}^{22}\lambda^{|b_Q^2-b_L^2|}+c_{QL}^{12}\frac{ V_{ud}}{V_{us}}\lambda^{|b_Q^1-b_L^2|}+c_{QL}^{32}\frac{V_{ub}}{V_{us}}\lambda^{|b_Q^3-b_L^2|}\right)     \,, \\
 - &\frac{2v^2}{\Lambda^2}  \frac{m_K^2}{m_\tau (m_u+m_{d})}c_{DE}^{23} \lambda^{ |b_D^2-b_E^3|}\left( c_{QL}^{13} \lambda^ {|b_Q^1-b_L^3|}+c_{QL}^{23}\frac{V_{us}}{V_{ud}} \lambda^{|b_Q^2-b_L^3|}+c_{QL}^{33}\frac{V_{ub}}{V_{ud}}\lambda^{|b_Q^3-b_L^3|}\right)\notag \\
  +&\frac{2v^2}{\Lambda^2}  \frac{m_K^2}{m_\mu (m_u+m_{d})}c_{DE}^{22} \lambda^{ |b_D^2-b_E^2|}\left( c_{QL}^{12} \lambda^ {|b_Q^1-b_L^2|}+c_{QL}^{22}\frac{V_{us}}{V_{ud}} \lambda^{|b_Q^2-b_L^2|}+c_{QL}^{32}\frac{V_{ub}}{V_{ud}}\lambda^{|b_Q^3-b_L^2|}\right)\,.\notag
\end{align}

\subsection{Constraints on FN charges} 
\label{sec:FN_constr}
The conditions discussed in \sec{sec:FN} 
on the FN charges can be further constrained by using the low-energy observables listed above. This procedure is clearly spurion-dependent, i.e., while the conditions of \sec{sec:FN} 
are linked to the SM and can be considered universal, the ones that we will derive below are associated to the specific extension of MFV. An important point to stress is that compliance with flavour tests does not lead to a single solution for the FN charges. Instead, there is a family of them. 

In order to reduce the parameter space, our strategy will be to use the following subset of processes to set constraints on the FN charges: (i) the universality ratios $R_{D^{(*)}}$, (ii) the global fits to $b\to s\ell\ell$ observables, together with (iii) precision tests of $Z\to\nu\bar\nu$, $\bar B_s\to\tau^{\mp}\mu^{\pm}$, $K_L\to \mu^\pm e^\mp$ and $\bar B_d\to\tau^-\mu^+$. This will leave us with a manageable number of potential solutions, which can then be analysed separately in the global fit.

Below we list the different processes. In order to set constraints on the FN charges, the phenomenological limits have been translated into powers of $\lambda$.

\begin{enumerate}
\item \boldmath$ Z\to\nu\bar\nu$\unboldmath. This decay happens in our setup dominantly through a top loop. The corresponding expression is listed in \eq{eq:rad_Znu_FN}. 
The experimental limits require $\left(\delta g^Z_{\nu_L}\right)_{\alpha\alpha}\leq \lambda^2$, which translates into
\begin{align}
  \left| b_L^\alpha \right| \geq 1\,.
\label{eq:FN_const1}
\end{align}

\item \boldmath$b\to s\mu^+\mu^-$\unboldmath. The magnitude of the coupling mediating the $b\to s\mu^+\mu^-$ transition can be inferred from the global fits of \cite{Alguero:2019ptt,Aebischer:2019mlg,Ciuchini:2019usw}. Among the possible scenarios, we require a sizeable left-handed contribution, with other structures suppressed. This means that from \eq{eq:bsmumu_FN} 
\begin{align}
\mathcal{C}_9^{2322}=&  - \mathcal{C}_{10}^{2322} \sim \lambda^2\,,
\end{align}
while all other Wilson coefficients listed in \eq{eq:bsmumu_FN} have additional suppressions. Counting the loop suppression in the SM compared to the tree-level leptoquark exchange as ${\cal O}(\lambda^2)$ leads to
  \begin{align}
 	 |b_L^2|+|b_L^2-2| = 4 \qquad {\Rightarrow} \qquad b_L^2 = -1 (+3) \,.
    	\label{eq:FN_const2}
 \end{align}
Suppression of the scalar and right-handed contributions means that
\begin{align}
	 |b_D^3-b_E^2|+|b_L^2-2| \geq 4\,, \qquad \text{and} \qquad  |b_L^2| + |b_D^2-b_E^2|  \geq  4 \,.
 	 \label{eq:FN_const3}
\end{align}
\item \boldmath$b\to c\tau^-\bar\nu$\unboldmath. The contribution to $b\to c\tau^- \nu$ is proportional to the Wilson coefficient $\mathcal{C}_L^{2333}$ defined in \eq{eq:btoc_coeffs1}, which consists of three contributions. Due to the above constraints, the leading contribution is given by $[\mathcal{C}^{(3)}_{lq}]^{2333}$, while  $[\mathcal{C}^{(3)}_{lq}]^{3333}$ and  $[\mathcal{C}^{(3)}_{lq}]^{1333}$ are suppressed. The current measurements of $R_{D^{(*)}}$ require $[\mathcal{C}^{(3)}_{lq}]^{2333}\sim \lambda^2$, which translates into
\begin{align}
	|b_L^3|+|b_L^3-2| = 2 \qquad{\Rightarrow} \qquad b_L^3 = +1 (+2)\,.
    	\label{eq:FN_const4}
\end{align}
where $b_L^3=0$ is excluded by \eq{eq:FN_const1}.

A sizeable scalar contribution to $b\to c\tau^-\bar\nu$ decays is known to improve the quality of fits \cite{Bordone:2017bld,Bordone:2018nbg,Cornella:2019hct,Fuentes-Martin:2019mun,Murgui:2019czp,Shi:2019gxi}. We require it to  be comparable with the left-handed one, i.e.
\begin{align}
	|2-b_L^3| + |b_D^3-b_E^3| = 2\,.
	\label{eq:FN_const5} 
 \end{align}
 \item \boldmath$\bar B_d \to \tau^-\mu^+$\unboldmath. In order to have a sufficient suppression of the scalar contribution in this channel, we would need 
\begin{align}
	|3-b_L^2| + |b_D^3-b_E^3| \geq 5  \,.
     	\label{eq:FN_const6}
\end{align}
A comparison with \eq{eq:FN_const5} then singles out $b_L^2=-1$ as the only viable charge.

\item \boldmath$\bar B_s\to \tau^{\pm}\mu^{\mp}$\unboldmath. This decay mode is sensitive to
the FN charges of the second and third generation of left-handed leptons. The current experimental 
 constraints require a suppression of at least ${\cal O}(\lambda^2)$. This bound is automatically fulfilled for $\bar B_s\to \tau^-\mu^+$. For $\bar B_s\to \tau^+\mu^-$ this requires instead
\begin{align}
|b_L^2| + |b_L^3-2| \geq 2  \,,
\end{align}
which, given $b_L^2=-1$, singles out $b_L^3=1$ as the only viable solution.

 \item \boldmath$K_L\to \mu^{\pm} e^{\mp}$\unboldmath. The experimental limits on these modes set constraints on $b_L^1$. Currently the bounds require at least an ${\cal O}(\lambda^8)$ suppression at the amplitude level. This translates into 
\begin{align}
	|3-b_L^2| + |2-b_L^1| &\gtrsim 8 \,,\\
 	|3-b_L^1| + |2-b_L^2] &\gtrsim 8  \,.
\label{eq:FN_const7}
\end{align}
Given $b_L^2=-1$, the values that saturate the bounds are $b_L^1 = -2 (+8)$, which we will take as our benchmark points.

\end{enumerate}
Considering all the above, the left-handed FN charges are then constrained to the values
\begin{equation}
\begin{array}{ll}
b_Q^1=3\,, &\qquad b_L^1=-2 (+8)\,,\\[3pt]
b_Q^2=2\,, &\qquad b_L^2=-1\,, \\[3pt]
b_Q^3=0\,, &\qquad b_L^3=+1\,.
\end{array}
\end{equation}
Concerning the right-handed FN charges,
with the previous constraints \eq{eq:FN_const5}
becomes
\begin{align}
	|b_D^3-b_E^3| = 1\,.
\end{align}
Using the information on the fermion Yukawa couplings, only the combinations 
\begin{equation}
(b_D^3,b_E^3)=\{(-3,-2), (3,4)\}
\end{equation}
get selected.
Adding the constraints of \eq{eq:FN_const3}, 
one is then left with the three following combinations for the second and third generations:
\begin{align}
 (b_D^2,b_D^3; \ b_E^2,b_E^3) = \{ (-3,-3; \ +4,-2) \,, \ (+7,-3;\ -6,-2)\,, \ (+7,+3;\ -6,+4) \}\,.
\end{align}

Considering the $2^3=8$ choices for $b_L^1,b_D^1,b_E^1$ together with the previous constraint, we are left with 24 potential solutions.
In order to determine which of these solutions can better reproduce the phenomenology of all the processes illustrated in \sec{sec:relevant}, in the next Section we will perform a fit. This will also allow us to carry a detailed examination of the individual phenomenological features of each solution.


\section{Fit results and discussion}
\label{sec:fit}

The fit is performed by the minimisation of the log likelihood, constructed as
\begin{equation}
\log(\mathcal{L}) = -\frac{1}{2}\sum_{i \, \in \,\text{obs} } \left(\mathcal{O}^i_{th}-\mathcal{O}^i_{exp}\right)^T \Sigma_i^{-1} \left(\mathcal{O}^i_{th}-\mathcal{O}^i_{exp}\right) \,,
\end{equation}
where  $\Sigma$ is the covariance matrix, and 
the sum runs over all the observables discussed  in \sec{sec:U1_model}. 
The statistical analysis is performed using the package MultiNest \cite{Buchner:2014nha}
for  each of the 24 FN charge assignments.
For the purpose of this paper we will not vary the CKM matrix elements and instead fix them to the UTFit NP fit \cite{UTFit}. This choice is motivated by the expectation that uncertainties on the CKM parameters will be negligible. All other inputs, e.g. masses and lifetimes, are taken from the PDG \cite{Tanabashi:2018oca}.
To reduce the number of free parameters in the fit, we will make a number of additional
simplifying assumptions, namely 
\begin{itemize}
\item A single (real) flavour-independent Wilson coefficient for each of the spurion entries up to a relative sign, i.e.
\begin{equation}
c_{QL}^{i\alpha}=\pm \, \mathcal{C}_{QL}\,, \quad \text{and} \quad  c_{DE}^{i\alpha}=\pm\,\mathcal{C}_{DE}\,,
\end{equation}
where the capital coefficients $\mathcal{C}_{QL}$ and $\mathcal{C}_{DE}$ are taken positive.
\item As observed in \sec{sec:simp}, the relation
\begin{align}
{\cal{C}}_{lq}^{(1)}=\,{\cal{C}}_{lq}^{(3)}
\end{align}
does not need to hold in a spurion approach and could in principle be relaxed. However, we will assume it to be in place. The main feature is that the processes $\tau\to 3\mu$, $B \to K^*\nu\bar\nu$, $K\to \pi\nu{\bar{\nu}}$ and $Z\to\tau\tau$ receive no corrections in that case (see the \app{app:2}). Note that this assumption is an intrinsic feature of the $U_1$ leptoquark model at tree level, which has been implicitly used in \sec{sec:U1_model}.
\item We set the value of the effective NP scale to be $\Lambda= 2 \, \text{TeV}$. 
\end{itemize}
All of these assumptions could be easily relaxed if more or more precise data becomes available from 
present and future experiments (for the prospects, see e.g.\ 
\cite{Bediaga:2012py,Kou:2018nap,Cerri:2018ypt}). For the sake of this work, we are merely interested 
in constraining the generic order of magnitude of the spurion entries. Of course, with the 
above assumptions we may actually miss FN scenarios with particularly tuned parameters.

Regarding the signs of the spurion entries, we already mentioned in \sec{sec:FCNCs} 
that a negative contribution to ${\cal{C}}^{2322}_9$ requires
\begin{equation}
c_{QL}^{22} c_{QL}^{32}<0 \,.
\end{equation} 
In what follows, we will choose $c_{QL}^{32}=-\,\mathcal{C}_{QL}<0$ and $c_{QL}^{22}=+\,\mathcal{C}_{QL}>0$. We have verified explicitly that the opposite choice does not significantly increase 
the $\chi^2$ value. All other entries $c_{QL}^{i\alpha}$ are chosen to be positive. 
One also needs
\begin{equation}
c_{DE}^{33} = -\,\mathcal{C}_{DE}< 0\,,
\end{equation}
in order to have a constructive interference between left-handed and scalar contributions in $R_{D^{(*)}}$. The sign of the other entries for $c_{DE}^{i\alpha}$ are not constrained by phenomenology and for simplicity we take them positive.
The complete list of the observables used for the fit is given in \sec{sec:U1_model}, 
and the corresponding experimental measurements and SM predictions can be found in \app{app:2}.
\begin{table}[h]
\begin{center}
\renewcommand{\arraystretch}{1.2} 
\begin{tabular}{c c c c c c c c c c c c c} 
\toprule
Scenario &  $b_L^1$ &$b_D^1$&$b_D^2$&$b_D^3$ & $b_E^1$ & $b_E^2$ & $b_E^3$ & $\mathcal{C}_{QL}$ & $\mathcal{C}_{DE}$ \\
\midrule
1a &\multirow{5}{*}{$-2$} & $10$ & $-3$ & $-3$ & $-11$ & $4$ & $-2$ & $1.10 \pm 0.07$ & $0.72 \pm 0.22$ \\
1b & & $10$ & $7$ & $-3$ & $-11$ & $-6$ & $-2$ & $1.07\pm0.08$ & $6.4\pm 1.8$\\
1c & & $10$ & $7$ & $3$ & $-11$ & $-6$ & $4$ & $1.07\pm 0.08$ & $7.2\pm 2.1$ \\
1d & & $-4$ & $-3$ & $-3$ & $-11$ & $4$ & $-2$ &  $1.10\pm 0.09$ & $0.74 \pm 0.28$ \\
1e & & $-4$ & $-3$ & $-3$ & $7$ & $4$ & $-2$ & $1.10\pm 0.09$ & $0.73 \pm 0.28$\\
\midrule
2a &\multirow{6}{*}{$+8$}  & $10$ & $-3$ & $-3$ & $17$ & $4$ & $-2$ & $1.10\pm 0.10$ & $0.74 \pm 0.26$ \\
2b & & $10$ & $7$ & $-3$ & $-1$ & $-6$ & $-2$ & $1.09\pm 0.09$ & $0.42 \pm 0.25$ \\
2c & & $10$ & $7$ & $-3$ & $17$ & $-6$ & $-2$ & $1.08\pm0.09$ & $4.6\pm 1.4$ \\
2d & & $10$ & $7$ & $3$ & $-1$ & $-6$ & $4$ & $1.07\pm0.10$ & $7.1\pm 2.0$ \\
2e & & $10$ & $7$ & $3$ & $17$ & $-6$ & $4$ & $1.08\pm0.09$& $4.8\pm1.3$\\
2f & &$-4$ & $-3$ & $-3$ & $17$ & $4$ & $-2$ &$1.10\pm0.09$ & $0.74\pm0.28$\\
\bottomrule
\end{tabular}
\caption{Viable solutions for FN charges identified from the fit.}
\label{tab:viables_sol}
\end{center}
\end{table}

With the above assumptions, out of the $24$ possible solutions, only $11$ provide an acceptable fit, 
i.e.\ they show a sizeable reduction of the $\chi^2$ value with respect to the SM one. The values of the charges associated with these solutions and the posterior for the Wilson coefficients are given in 
\Table{tab:viables_sol}. For reference, the corresponding FN scaling of the flavour spurions together with the fermionic rotation matrices from the FN basis to the mass eigenbasis are listed in \app{app:scaling}. We have explicitly checked that the 11 solutions (i) generate a $b\to s\mu^+\mu^-$ transition dominated by the left-handed operators only, (ii) do not violate the existing bounds on the $B_c$ lifetime, and (iii) do not produce sizeable contributions to $b\to s e^+ e^-$.

As it can be seen in \Table{tab:viables_sol}, 
the values of $\mathcal{C}_{QL}$ are rather uniform, which is due to the constraints from $b\to s\mu^+\mu^-$ data. The values of $\mathcal{C}_{DE}$ instead roughly fall into two groups: for half the solutions it is rather large (between 4 and 7), whereas for the other half it is one order of magnitude smaller. We have checked that the solutions with a large scalar Wilson coefficient have FN charges that suppress the scalar contribution to the $\bar B_s\to \tau\tau$ and $\bar B_s\to \tau\mu$ decays. In turn, the solutions with small scalar Wilson coefficients are $\lambda$-enhanced. An exception is the solution 2d, which has a small scalar coefficient to compensate for a large power-counting factor in $\bar B_s\to\mu^+e^-$. Despite these differences, we note that the values of all the Wilson coefficients are in the ballpark of what one would expect from power-counting arguments.

An interesting point to remark is that all the solutions have roughly the same quality of fit. The global $\chi^2$ for each viable solution improves the SM one by approximately a factor 3. Further improvements are however blocked by the tension between the $W$ universality tests and $R_{D^{(*)}}$, which is actually driving the fit. This tension is not specific of our setup but rather a generic feature of any EFT involving $\tau$ decays.  Further experimental measurements are needed to clarify this interesting issue.

\begin{figure}\begin{center}
\centering
\includegraphics[scale=0.8]{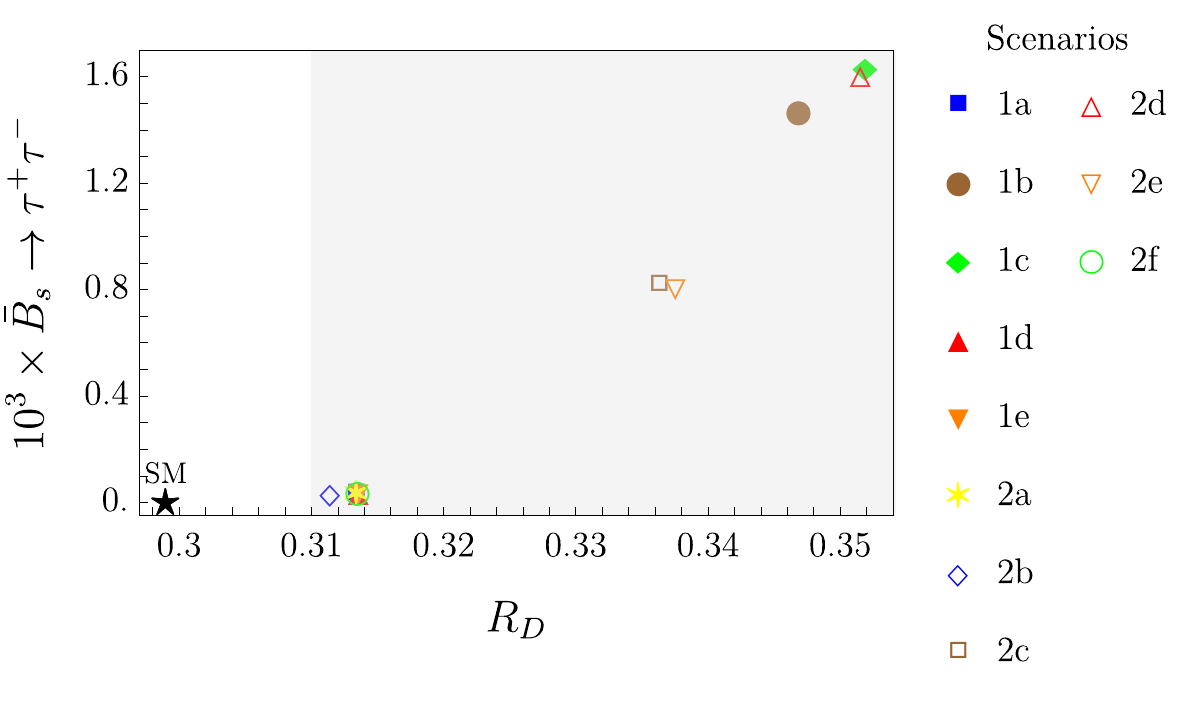}
\end{center}
\caption{Correlation between $\bar B_s\to\tau^+\tau^-$ and $R_{D}$ for all the $11$ scenarios listed in \Table{tab:viables_sol}. 
The grey band represents the $1\sigma$ region for $R_D$ (see \Table{tab:LFU_btoc} for the numerical values used in the plot). 
}
\label{fig:BstautauvsRD}
\end{figure}

Despite leading to similar $\chi^2$ values, each solution has a number of distinctive features, in the form of different predictions for the various observables. An example is showed in \fig{fig:BstautauvsRD}, where the combined prediction for $\bar B_s\to \tau^+\tau^-$ and $R_D$ is illustrated for each scenario. These two observables happen to be directly correlated: the 5 solutions with large scalar Wilson coefficients give a larger value for $R_D$, as expected, and also predictions for $\bar B_s\to \tau^+\tau^-$ close to the present experimental limit. The other 6 solutions are clustered close to the SM value. As already pointed out in other works, e.g. in \cite{Cornella:2019hct}, a future investigation of $\bar B_s\to \tau^+\tau^-$ is an important and complementary observable in the study of the $B$ anomalies. 

Note that all scenarios predict a value for $R_D$ within $1\sigma$ from the experimental result. Instead we find that our predictions for $R_{D^*}$ hover in the $1.5-2\sigma$ range. This difference between the NP effects on $R_D$ and $R_{D^*}$ lies in the relative weight of the scalar contributions, which enhance the former but not the latter.

\begin{figure}[h]
\begin{center}
\includegraphics[scale=0.4]{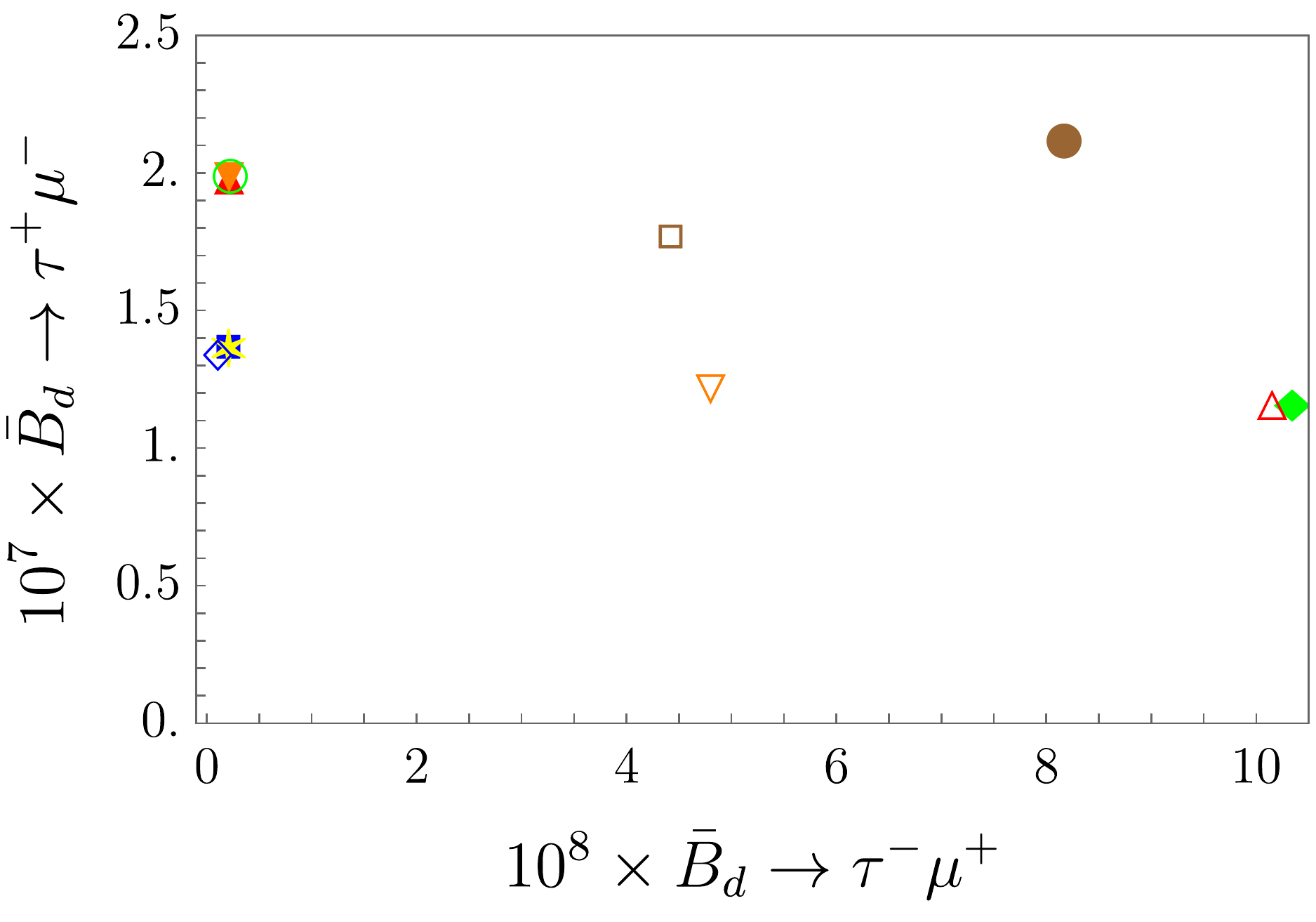}
\hspace{5mm}
\includegraphics[scale=0.4]{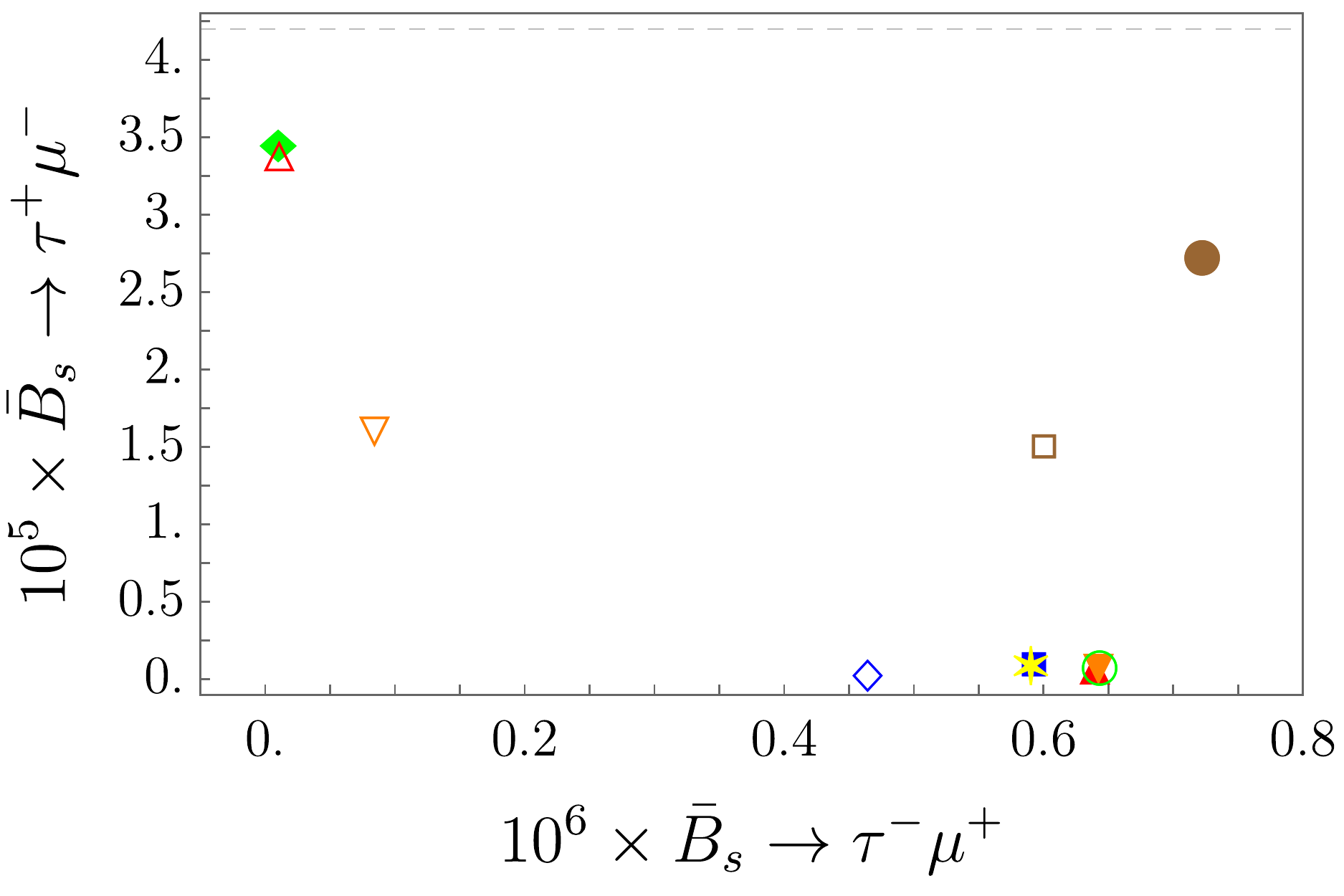}
\end{center}
\caption{Correlation between the $\bar B_{d,s}\to \tau^-\mu^+$ and $\bar B_{d,s}\to \tau^+\mu^-$ modes for each scenario. The different solutions are labeled as in \fig{fig:BstautauvsRD}. On the left panel, the experimental limit is at $\mathcal{O}(10^{-6})$ and accordingly is not displayed. On the right panel, the grey dashed line is the current LHCb upper limit \cite{Aaij:2019okb}. This includes both  $\bar B_{s}\to \tau^-\mu^+$ and $\bar B_{s}\to \tau^+\mu^-$ modes, so what is shown is a conservative upper bound.
}
\label{fig:LFV_Bdecay}
\end{figure}

Another interesting point to stress is that with leptoquark spurions one expects asymmetric predictions for 
decay modes with charge-conjugated final states. As illustrative examples, in \fig{fig:LFV_Bdecay}
we display the predictions for $\bar B_{s,d}\to\tau^\pm\mu^\mp$ for the different solutions. These modes are particularly sensitive to the scalar contributions via a chiral enhancement. Note that our solutions consistently give larger predictions for the LFV decays $\bar B_{s,d}\to\tau^-\ell^+$ over $\bar B_{s,d}\to\tau^+\ell^-$ by orders of magnitude. This feature of our setup indicates that separate analyses for the different modes are very informative. The current experimental bounds for $\bar B_s\to \tau^\pm\mu^\mp$ and $\bar B_d\to \tau^\pm\mu^\mp$ instead  comprise the sum of the charge conjugated modes \cite{Aaij:2019okb}. Experimentally, separate analyses are hindered by the need to distinguish $B$ from $\bar{B}$ in the initial state. Both Belle II and LHCb can perform this tagging, but the penalty in terms of statistics is significant, especially for the LHCb \cite{Bediaga:2018lhg}. This suggests to first do an untagged analysis, as is done currently, in order to search for a signal in these LFV $B$ decays. If a signal is detected, the next step would be to tag the initial state and assess if one of the final states, $\tau^-\mu^+$ or $\tau^+\mu^-$, dominates.

\section{Conclusions}

\label{sec:conclusions}

In this paper we have suggested an EFT-based framework to study flavour processes in the presence of new physics. Our starting point is a general description of the possible bosonic couplings to Standard Model Dirac bilinears in terms of flavour spurions, which represents a natural generalisation of minimal flavour violation. 
The hierarchies among the new flavour coefficients are fixed by a power-counting scheme, which is
defined in terms of generalised Froggatt-Nielsen charge assignments for SM quarks and leptons.  This 
has the theoretical advantage of being transparent, model-independent 
and self-consistent with respect to higher-order corrections. 

In order to illustrate how such a framework can be used in practice, we have selected the spurions $\Delta_{QL}$ 
and $\Delta_{DE}$, motivated by a simplified $U_1$ leptoquark model. 
This represents a minimal extension of minimal flavour violation able to accommodate the present flavour anomalies.
Constraints on the FN-charge parameter space have been imposed both from the SM Yukawa structure and experimental limits on low-energy observables. This procedure narrows down the potential solutions for FN charges to $24$, which are then studied in detail through a rather exhaustive fit to low-energy flavour observables. The fit eventually selects $11$ phenomenologically viable solutions, which could be further distinguished if more precise data on leptonic $B$-meson decays becomes available in the future.  

Rather generically, the fit is dominated by the tension between constraints 
on lepton-flavour non-universal effects in the couplings of the $W$, the ratios $R_{D^{(*)}}$ 
and deviations of $\bar B_s\to\tau^+\tau^-$ from the SM prediction. 
This suggests that, if the anomalies in $R_{D^{(*)}}$ persist, a reanalysis of LFU in the $W$ couplings is necessary to better understand their correlation. 
More generally, this rather generic tension is a strong motivation to improve the measurements on processes involving the tau lepton either in the initial or final state. This is already one of the goals at LHCb.

Another interesting point to stress is that spurions associated with leptoquarks have the potential to generate asymmetric predictions for leptonic decay modes of neutral $B$-mesons with charge-conjugated final states. Specifically, we have pointed out that all our solutions lead to asymmetries in the prediction of the LFV leptonic decays $\bar B_{s,d}\to \tau^\pm\mu^\mp$. As a result, it is important to carry out separate measurements of the decays $\bar B_{s,d}\to \tau^-\mu^+$ and $\bar B_{s,d}\to\tau^+\mu^-$. This requires tagging the initial state, which both LHCb and Belle II can do. However, since tagging efficiencies are rather low, these separate analyses should be done once a signal has been detected or a very high statistics is collected.

From the theoretical perspective it would be interesting to explore whether the identified solutions for FN charges could be understood from a specific dynamical mechanism in the context of some UV complete models, potentially embedded in some GUT scenario. In this regard recent work in \cite{Fornal:2018dqn,Smolkovic:2019jow} could be helpful.


\section*{Acknowledgements}
We are grateful to Giulio Dujany, Javier Fuentes-Mart\'{i}n and Diego Mart\'inez Santos for helpful discussions. 
M.B.\ and O.C.\ acknowledge support from 
the Deutsche Forschungsgemeinschaft (DFG, German Research Foundation) 
within the Research Unit FOR 1873 (“Quark Flavour Physics and Effective Field Theories”).
The research of T.F.\ is supported by the DFG
within the Collaborative Research Center  TRR 257
(``Particle Physics Phenomenology after the Higgs Discovery'').


\appendix
\section{From the SMEFT to the LEFT}
\label{app:A}
At energy scales relevant for flavour processes it is convenient to work with 
a low-energy effective theory (LEFT) where the heavy SM particles ($W$, $Z$, $h$ and $t$) have been integrated out
and electroweak symmetry breaking is manifest. For the processes we are considering in this paper, the tree level matching between both theories is straightforward.

The effects of the $U_1$ model considered in the main text at the electroweak scale can be described with the effective Lagrangian:
\begin{equation}
\begin{aligned}
\mathcal{L}_\text{eff}=\mathcal{L}_\text{SM}-\bigg\{ &[\mathcal{C}^{(3)}_{lq}]^{ij\alpha\beta} (\bar{Q}^i\gamma^\mu \sigma^a Q^j)(\bar{L}^\alpha\gamma_\mu \sigma^a L^\beta)+[\mathcal{C}^{(1)}_{lq}]^{ij\alpha\beta} (\bar{Q}^i\gamma^\mu Q^j)(\bar{L}^\alpha\gamma_\mu L^\beta) \\
+& [\mathcal{C}_{ed}]^{ij\alpha\beta}(\bar{d}^i\gamma^\mu d^j)(\bar{e}^\alpha\gamma_\mu e^\beta) 
+[\mathcal{C}_{ledq}]^{ij\alpha\beta}(\bar{Q}^i d^j)(\bar{e}^\alpha L^\beta)\bigg\} \,,
\label{eq:SMEFT_lagr_U1}
\end{aligned}
\end{equation}
where $\epsilon$ is the $2\times 2$ antisymmetric tensor, defined as $\epsilon_{01}=+1$.

The corresponding operators of LEFT (at tree level) can be found simply by going to the broken phase. 
To fix our conventions, we choose to work in the down-quark mass eigenbasis 
for the left-handed quark doublets. 
Regarding leptons, we adopt the charged-lepton mass eigenbasis and neglect 
neutrino mixing. With these conventions,
\begin{equation}
Q^i\equiv \begin{pmatrix}
V^*_{ij} u^j \\
 d^i
\end{pmatrix}
\,,
\qquad
L^\alpha \equiv
\begin{pmatrix}
 \nu^\alpha \\
e^\alpha
\end{pmatrix}\,.
\label{eq:SU_basis}
\end{equation}
In the broken phase the previous Lagrangian can thus be recast as
\begin{align}
\mathcal{L}_\text{eff}-\mathcal{L}_\text{SM} =-\frac{1}{\Lambda^2}\bigg\{&([\mathcal{C}^{(3)}_{lq}]^{ij\alpha\beta}+[\mathcal{C}^{(1)}_{lq}]^{ij\alpha\beta})[(\bar{u}^i\gamma^\mu P_L u^j)(\bar{\nu}^\alpha\gamma_\mu P_L \nu^\beta)+(\bar{d}^i\gamma^\mu P_L d^j)(\bar{e}^\alpha\gamma_\mu P_L e^\beta) ]]\notag\\
+&([\mathcal{C}^{(1)}_{lq}]^{ij\alpha\beta}-[\mathcal{C}^{(3)}_{lq}]^{ij\alpha\beta})[(\bar{u}^i\gamma^\mu P_L u^j)(\bar{e}^\alpha\gamma_\mu P_L e^\beta)+(\bar{d}^i\gamma^\mu P_L d^j)(\bar{\nu}^\alpha\gamma_\mu P_L \nu^\beta) ] \notag \\
+&2\,[\mathcal{C}^{(3)}_{lq}]^{ij\alpha\beta} \left[V^*_{mj}(\bar{d}^i\gamma^\mu P_L u^m)(\bar{\nu}^\alpha\gamma_\mu P_L e^\beta)+V_{mi}(\bar{u}^m\gamma^\mu P_L d^j)(\bar{e}^\alpha\gamma_\mu P_L \nu^\beta)\right] \notag \\
+&[\mathcal{C}_{leqd}]^{ij\alpha\beta} \left[V_{mi}(\bar{u}^m P_R d^j)(\bar{e}^\alpha P_L \nu^\beta)+(\bar{d}^i P_R d^j)(\bar{e}^\alpha P_L e^\beta)\right] \notag\\
+&[\mathcal{C}_{ed}]^{ij\alpha\beta}(\bar{d}^i\gamma^\mu P_R d^j)(\bar{e}^\alpha\gamma_\mu P_R e^\beta) 
\bigg\}\,,
\label{eq:dim6U1broken}
\end{align}
where $P_{L(R)}$ is the projector on the left(right)-handed components of fermion fields.
\section{Observables}
\label{app:2}
\subsection{$d_j\to d_i \ell_\alpha\ell_\beta$}
The effective Lagrangian describing a generic $d_j\to d_i \ell_\alpha\ell_\beta$ FCNC transition reads
\be
\begin{aligned}
\mathcal{L}_\text{NP}(d_j\to d_i \ell_\alpha\ell_\beta)=\frac{4 G_F}{\sqrt{2}}\frac{\alpha_\text{EM}}{4\pi}V_{td_j}V^*_{td_i}&\left[(\mathcal{C}_9^\text{SM}\delta_{\alpha\beta}+\mathcal{C}_9^{ij\alpha\beta})\mathcal{O}_9^{ij\alpha\beta}+(\mathcal{C}_{10}^\text{SM}\delta_{\alpha\beta}+\mathcal{C}_{10}^{ij\alpha\beta})\mathcal{O}_{10}^{ij\alpha\beta}\right. \\
&+\mathcal{C}_9^{\prime\, ij\alpha\beta}\mathcal{O}^{\prime\, ij\alpha\beta}_{9}
+\mathcal{C}_{10}^{\prime\, ij\alpha\beta}\mathcal{O}^{\prime\, ij\alpha\beta}_{10}+\mathcal{C}_S^{ ij\alpha\beta}\mathcal{O}^{ ij\alpha\beta}_S \\
&\left.+\mathcal{C}_P^{ ij\alpha\beta}\mathcal{O}^{ ij\alpha\beta}_P +\mathcal{C}_S^{\prime\, ij\alpha\beta}\mathcal{O}^{\prime\, ij\alpha\beta}_S+\mathcal{C}_P^{\prime\, ij\alpha\beta}\mathcal{O}^{\prime\, ij\alpha\beta}_P \right ]\,,
\end{aligned}
\label{eq:lagrangian_bsll}
\ee
where
\begin{align}
\mathcal{O}_9^{ij\alpha\beta} = & \,(\bar{d}_i \gamma^\mu P_L d_j)(\bar{\ell}_\beta \gamma_\mu \ell_\alpha)  \,, & \mathcal{O}_{10}^{ij\alpha\beta} = &\,(\bar{d}_i \gamma^\mu P_L d_j)(\bar{\ell}_\beta \gamma_\mu\gamma_5 \ell_\alpha) \,, \\
\mathcal{O}^{\prime\, ij\alpha\beta}_9 = &  \,(\bar{d}_i \gamma^\mu P_R d_j)(\bar{\ell}_\beta \gamma_\mu \ell_\alpha)\,, & \mathcal{O}^{\prime\, ij\alpha\beta}_{10} = & \,(\bar{d}_i \gamma^\mu P_R d_j)(\bar{\ell}_\beta \gamma_\mu\gamma_5 \ell_\alpha)\,, \\
\mathcal{O}_S^{ij\alpha\beta} = &  \,(\bar{d}_iP_R d_j)(\bar{\ell}_\beta  \ell_\alpha) \,,
& \mathcal{O}_{P}^{ij\alpha\beta} = &\,(\bar{d}_i  P_R d_j)(\bar{\ell}_\beta \gamma_5 \ell_\alpha) \,, \\
\mathcal{O}_S^{\prime\, ij\alpha\beta} = &  \,(\bar{d}_iP_L d_j)(\bar{\ell}_\beta  \ell_\alpha) \,,
& \mathcal{O}_{P}^{\prime\, ij\alpha\beta} = &\,(\bar{d}_i  P_L d_j)(\bar{\ell}_\beta \gamma_5 \ell_\alpha) \,.
\label{eq:base_NC}
\end{align}
Using the results of the previous Appendix, the matching to SMEFT Wilson coefficients reads
\be
\begin{aligned}
\mathcal{C}_{9}^{ij\alpha\beta} = - \,\mathcal{C}_{10}^{ij\alpha\beta} =& +\frac{v^2}{\Lambda^2}\frac{\pi}{\alpha_\text{EM} |V_{tb}V^*_{ts}|} \left([\mathcal{C}_{lq}^{(3)}]^{ij\alpha\beta}+[\mathcal{C}_{lq}^{(1)}]^{ij\alpha\beta}\right)\,, \\
\mathcal{C}_{9}^{\prime\, ij\alpha\beta} = +\, \mathcal{C}_{10}^{\prime\, ij\alpha\beta} = &+ \frac{v^2}{\Lambda^2}\frac{\pi}{\alpha_\text{EM} |V_{tb}V^*_{ts}|}  [\mathcal{C}_{l d}]^{ij\alpha\beta}\,, \\
\mathcal{C}_S^{ij\alpha\beta} = -\,\mathcal{C}_{P}^{ij\alpha\beta} = & +\frac{v^2}{\Lambda^2}\frac{\pi}{\alpha_\text{EM} |V_{tb}V^*_{ts}|} \,[\mathcal{C}_{leqd}]^{ij\alpha\beta}\,, \\
\mathcal{C}_S^{\prime\, ij\alpha\beta} = +\,\mathcal{C}_{P}^{\prime\, ij\alpha\beta} = & + \frac{v^2}{\Lambda^2}\frac{\pi}{\alpha_\text{EM} |V_{tb}V^*_{ts}|} \,[\mathcal{C}_{leqd}^*]^{ji\beta\alpha}\,.
\end{aligned}
\label{eq:observables_bsll}
\ee
From the channels considered in the main text, the most stringent bounds come from the $B\to K^{(*)}\ell\ell$ decays. We chose to constrain the NP Wilson coefficients with the output of global fits \cite{Alguero:2019ptt,Aebischer:2019mlg,Ciuchini:2019usw}. \\

Constraints from FCNCs also come from two-body leptonic decays, including LFV modes. The decay rate of a generic meson $P_{ij}= d^j \bar{d}^i$ into a lepton pair $\bar{\ell}_\alpha \ell_\beta$ generated by \eq{eq:lagrangian_bsll} reads
\begin{equation}
\begin{aligned}
\mathcal{B}(P_{ij}\to\ell^-_\alpha \ell^+_\beta)=&\frac{\tau_P}{64\pi^3}\frac{\alpha_\text{EM}^2G_F^2}{m_P^3} f_P^2 \, \lambda^{1/2}(m_P^2,m_\alpha^2,m_\beta^2) \times \\
\times&\left\{[m_P^2-(m_{\ell_\alpha}-m_{\ell_\beta})^2]\left|(m_{\ell_\alpha}+m_{\ell_\beta})(\mathcal{C}^{ij\alpha\beta}_{10}-\mathcal{C}^{\prime\, ij\alpha\beta}_{10})+\frac{m_P^2}{m_i+m_j}\mathcal{C}^{ij\alpha\beta}_P\right|^2\right. \\
&+\left.[m_P^2-(m_{\ell_\alpha}+m_{\ell_\beta})^2]\left|(m_{\ell_\alpha}-m_{\ell_\beta})(\mathcal{C}^{ij\alpha\beta}_{9}-\mathcal{C}^{\prime\, ij\alpha\beta}_{9})+\frac{m_P^2}{m_i+m_j}\mathcal{C}^{ij\alpha\beta}_S\right|^2\right\} \,.
\end{aligned}
\end{equation}
The full list of modes, together with their experimental bounds, is displayed in Table~\ref{tab:LFV_FCNC}.

\begin{table}
\begin{center}
\renewcommand{\arraystretch}{1.2} 
\begin{tabular}{c c }
\toprule
Observable & Upper limit  \\
\midrule
$\bar B_d\to\tau^\pm\mu^\mp$ & $1.4\cdot 10^{-5}$ \cite{Aaij:2019okb}  \\ 
$\bar B_d\to\tau^- e^+$ & $2.8 \cdot 10^{-5}$ \cite{Aubert:2008cu}  \\ 
$\bar B_d\to\mu^\pm e^\mp$ & $3.7 \cdot 10^{-9}$ \cite{Aaij:2013cby} \\ 
$\bar B_s\to\mu^\pm e^\mp$ & $1.4 \cdot 10^{-9}$ \cite{Aaij:2013cby} \\ 
$K_L\to\mu^\pm e^\mp$ & $4.7\cdot 10^{-12}$ \cite{Ambrose:1998us}  \\ 
$\bar B_s\to \tau^\pm\mu^\mp$ & $4.2 \cdot 10^{-5}$ \cite{Aaij:2019okb}\\
$\bar B_s\to \tau^+\tau^-$ & $6.8 \cdot 10^{-3}$ \cite{Aaij:2017xqt}\\
\toprule
\end{tabular}
\caption{Experimental measurements of semileptonic LFV decays.}
\label{tab:LFV_FCNC}
\end{center}
\end{table}


\subsection{$d_j\to u_i \ell_\alpha\bar\nu_\beta$}
The charged-current transitions, $d_j\to u_i \ell_\alpha\bar\nu_\beta$, are described by the following Lagrangian
\be
\label{eq:btoc}
\begin{aligned}
\mathcal{L}(d_j\to u_i \ell_\alpha \bar\nu_\beta)=-\frac{4G_F}{\sqrt{2}} V_{ij} &\left[(\delta_{\alpha\beta}+\mathcal{C}_L^{ij\alpha\beta})(\bar{u}^i\gamma^\mu P_L d^j)(\bar{e}^\alpha\gamma_\mu P_L \nu^\beta)+ \mathcal{C}_{S}^{ij\alpha\beta}(\bar{u}^i P_Rd^j)(\bar{e}^\alpha P_L \nu^\beta)\right ] \,.
\end{aligned}
\ee
The NP Wilson coefficients read
\begin{align}
\mathcal{C}_{L}^{ij\alpha\beta}=&+ \frac{v^2}{\Lambda^2}\, \sum_{m=1}^3 \frac{V_{im}}{V_{ij}} [\mathcal{C}^{(3)}_{lq}]^{mj\alpha\beta}  \,,  \label{eq:btoc_coeffs1}\\
\mathcal{C}_{S}^{ij\alpha\beta}=&+\frac{v^2}{2\Lambda^2}\,  \sum_{m= 1}^3 \frac{V_{im}}{V_{ij}} [\mathcal{C}_{leqd}]^{mj\alpha\beta}\,. \label{eq:btoc_coeffs2} 
\end{align}
The most interesting channels for testing the $b\to c$ transitions are $B$ meson decays. The observables driving the NP effects are the universality ratios $R_{D^{(*)}}$. Their expressions in our framework are
\begin{equation}
\label{eq:observable: RDst}
\begin{aligned}
R_{D^{(*)}}=& R_{D^{(*)}}^{\text{SM}}  \frac{\left\vert 1+\, \mathcal{C}_L^{2333}+ \mathcal{F}^S_{D^{(*)}}(\tau)\, \mathcal{C}_{S}^{2333} \right\vert^2+\sum_{\beta \neq 3}\left\vert\mathcal{C}_L^{233\beta} +\mathcal{F}^S_{D^{(*)}}(\tau)\mathcal{C}_S^{233\beta} \right\vert^2}{\left\vert1+ \, \mathcal{C}_{L}^{2322}+\mathcal{F}^S_{D^{(*)}}(\mu) \mathcal{C}_{S}^{2322}\right\vert^2+\sum_{\beta \neq 2}\left\vert\mathcal{C}_L^{232\beta}+ \mathcal{F}^S_{D^{(*)}}(\mu)\mathcal{C}_S^{232\beta} \right\vert^2} \,,
\end{aligned}
\end{equation}
where the functions $\mathcal{F}^S_{D^{(*)}}(\ell)$ are placeholders for the integrals over kinematics and form factors associated with the scalar contributions for a $D$ or a $D^*$ and a charged lepton $\ell$. The numerical values for their linearised forms are given in the main text.

To understand if and how well universality holds for decays into light leptons, one can compare $|V_{cb}|$ as extracted from electron and muon modes. If we define as $|\tilde{V}_{cb}^\ell|$ the effective $|V_{cb}|$ in the presence of NP contributions associated with a lepton $\ell$, the universality in $\mu$ vs $e$ mode is measured by
\be
\frac{|\tilde{V}^e_{cb}|}{|\tilde{V}^\mu_{cb}|}=\left[ \frac{\vert1+\mathcal{C}_L^{2311}\vert^2+|\mathcal{C}_L^{2321}|^2+\vert\mathcal{C}_L^{2331}|^2}{\vert1+\mathcal{C}_L^{2322}\vert^2+|\mathcal{C}_L^{2312}|^2+|\mathcal{C}_L^{2332}|^2}\right]^{\frac{1}{2}} \,.
\ee
Contributions from scalar operators are suppressed by the lepton mass and can be safely neglected. \\

Finally, it is also interesting to consider the leptonic decay modes of charged $B_q$ mesons, where $q=u,c$. The corresponding branching ratio reads
\begin{align}
\mathcal{B}(B_q\to \ell\bar\nu) &=\mathcal{B}(B_q\to \ell\bar\nu)\vert_\text{SM}\left(\left|1+\mathcal{C}_L^{q3\ell\ell}+\frac{m_{B_q}^{2}}{m_\ell (m_b+m_q)}\mathcal{C}_{S}^{q3\ell\ell}\right|^2  \right.
\cr  & \qquad \phantom{\mathcal{B}(B_q\to \ell\bar\nu)\vert_\text{SM}} \left. +
\sum_{\ell\neq\ell^\prime}\left|\mathcal{C}_L^{q3\ell\ell^\prime}+\frac{m_{B_q}^{2}}{m_\ell (m_b+m_q)}\mathcal{C}_{S}^{q3\ell\ell^\prime}\right|^2 \right) \,.
\end{align}

\begin{table}
\begin{center}
\renewcommand{\arraystretch}{1.2} 
\begin{tabular}{c c c c}
\toprule
Observable & Measurement & Correlation & SM \\
\midrule
$R_D$	& $0.340 \pm0.027 \pm 0.013$ \cite{Amhis:2019ckw}	&\multirow{2}{*}{-0.38} &  $0.299\pm 0.003$ \cite{Bigi:2016mdz,Bernlochner:2017jka,Jaiswal:2017rve,Amhis:2019ckw} \\
$R_{D^*}$ & $0.295 \pm 0.011 \pm 0.008 $ \cite{Amhis:2019ckw}	&		& $0.258\pm0.005$ \cite{Bigi:2017jbd,Bernlochner:2017jka,Jaiswal:2017rve,Amhis:2019ckw}\\
\midrule
$V_{cb}\vert_D$ & $1.004(42)$ \cite{Jung:2018lfu}& \multirow{2}{*}{-} & 1. \\
$V_{cb}\vert_{D^*}$ & $0.97(4)$ \cite{Jung:2018lfu} &  &1.\\
\toprule
\end{tabular}
\caption{Experimental measurements, SM predictions and correlations for $b\to c$ transitions.}
\label{tab:LFU_btoc}
\end{center}
\end{table}

\subsection{$B\to K^{(*)}\nu\bar\nu$}
In $b\to s\nu\bar\nu$ transitions, the only NP contribution comes from left-handed operators. The relevant Lagrangian is 
\be
\begin{aligned}
\mathcal{L}(b\to s\nu\bar\nu)= +\frac{4G_F}{\sqrt{2}}\frac{\alpha_\text{EM}}{4\pi}V_{tb}V^*_{ts}&\left[\left(C_\nu^\text{SM}+\mathcal{C}_{L}^{33}\right)(\bar{s}\gamma^\mu P_L b)(\bar{\nu}^{\tau}\gamma_\mu P_L \nu^{\tau}) \right. \\
&+\left.\mathcal{C}_{L}^{3\alpha} (\bar{s}\gamma^\mu P_L b)(\bar{\nu}^{\tau}\gamma_\mu P_L\nu^\alpha)
+\mathcal{C}_{L}^{\alpha 3} (\bar{s}\gamma^\mu P_L b)(\bar{\nu}^\alpha \gamma_\mu P_L \nu^\tau) \right] \,,
\end{aligned}
\label{eq:lagrangian_bsnunu}
\ee
where $C_\nu^\text{SM}\approx -6.35$ \cite{Buras:2014fpa}. The remaining NP Wilson coefficients are generically given by
\begin{align}
\mathcal{C}_{L}^{\alpha\beta} = & \,+\frac{v^2}{\Lambda^2}\frac{\pi}{\alpha_\text{EM}|V_{tb}V^*_{ts}|}\left([\mathcal{C}_{lq}^{(1)}]^{23\alpha\beta}-[\mathcal{C}_{lq}^{(3)}]^{23\alpha\beta}\right) \,.
\end{align}
As it can be seen from \eq{eq:lagrangian_bsnunu}, final states with different neutrino species do not interfere with the SM contribution and are heavily suppressed. We will neglect them in the following.


\subsection{Modification of leptonic $Z$ couplings}
The $Z$ couplings with left-handed $\tau$ leptons and neutrinos are modified by one-loop corrections due to the RG evolution of semileptonic operators, mostly through the top loop contribution. The leading log results have been computed in \cite{Feruglio:2017rjo} and correspond to
\begin{align}
\left(\delta g^Z_{\nu_L}\right)_{\alpha\beta} =& -\frac{1}{8\pi^2}\frac{v^2}{\Lambda^2} 3 y_t^2 \left([\mathcal{C}^{(1)}_{lq}]^{33\alpha\beta}+[\mathcal{C}^{(3)}_{lq}]^{33\alpha\beta}\right)L_t \,,\label{eq:rad_Znu} \\
\left(\delta g^Z_{\tau_L}\right)_{\alpha\beta} =& -\frac{1}{8\pi^2}\frac{v^2}{\Lambda^2} 3 y_t^2 \left([\mathcal{C}^{(1)}_{lq}]^{33\alpha\beta}-[\mathcal{C}^{(3)}_{lq}]^{33\alpha\beta}\right)L_t \,, 
\end{align}
where $L_{t}= \log(\Lambda / m_{t})$ and $\alpha,\beta$ are the flavour indices of the final-state leptons. \\
\subsection{LFU in $W$ vertices}

$W$ vertices with leptons have been tested at the permille level and constrain the structure of any NP scenario. We can study these constraints via both leptonic and hadronic $\tau$ decays. 

For leptonic decays, we will define the relevant Lagrangian as
\begin{equation}
\mathcal{L}(e^\alpha \to e^k \nu^\beta \bar\nu^k)=- \frac{4 G_F}{\sqrt{2}} \left[\delta_{\alpha\beta}-(c_t^{cc})^{\alpha\beta}\right] (\bar{e}^\alpha \gamma_\mu P_L \nu^\beta)(\bar{\nu}^k\gamma^\mu P_L e^k) + \text{h.c.} \,,
\end{equation}
where 
\begin{equation}
(c_t^{cc})^{\alpha\beta}= \frac{v^2}{\Lambda^2}\frac{3 y_t^2}{16\pi^2}\left[\mathcal{C}_{lq}^{(3)}\right]^{33\alpha\beta}\log\left(\frac{\Lambda^2}{m_t^2}\right)
\end{equation}
takes into account the (leading) top contribution. 

In order to analyze the deviations in the $W$ couplings to the different leptons, it is common to define the following ratios
\be
\left(\frac{g_\tau}{g_\mu}\right)_\text{lept}= \left[\frac{\mathcal{B}(\tau\to e \nu\bar\nu)_\text{exp}/\mathcal{B}(\tau\to e \nu\bar\nu)_\text{SM}}{\mathcal{B}(\mu\to e \nu\bar\nu)_\text{exp}/\mathcal{B}(\mu\to e \nu\bar\nu)_\text{SM}}\right]^\frac{1}{2} \, ,
\label{eq:LFU_W}
\ee
which tests tau and muon universality. Similar definitions hold for other lepton combinations. 

In our framework, and using the results of \cite{Feruglio:2017rjo}, we obtain 
\be
\begin{aligned}
\left(\frac{g_\tau}{g_\mu}\right)_\text{lept}=& \left[\frac{\left\vert1-(c_t^{cc})^{33}\right\vert^2 + \left\vert (c_t^{cc})^{32} \right\vert^2+ \left\vert (c_t^{cc})^{31} \right\vert^2}{\left\vert1- (c_t^{cc})^{22}\right\vert^2 + \left\vert (c_t^{cc})^{23} \right\vert^2+ \left\vert (c_t^{cc})^{21} \right\vert^2}\right]^{1/2} \,,  \\
\left(\frac{g_\tau}{g_e}\right)_\text{lept}=& \left[\frac{\left\vert1- (c_t^{cc})^{33}\right\vert^2 + \left\vert (c_t^{cc})^{32} \right\vert^2+ \left\vert (c_t^{cc})^{31} \right\vert^2}{\left\vert1- (c_t^{cc})^{11}\right\vert^2 + \left\vert (c_t^{cc})^{12} \right\vert^2+ \left\vert (c_t^{cc})^{13} \right\vert^2}\right]^{1/2} \,,  \\
\left(\frac{g_\mu}{g_e}\right)_\text{lept}=&  \left[\frac{\left\vert1- (c_t^{cc})^{22}\right\vert^2 + \left\vert (c_t^{cc})^{21} \right\vert^2+ \left\vert (c_t^{cc})^{23} \right\vert^2}{\left\vert1- (c_t^{cc})^{11}\right\vert^2 + \left\vert (c_t^{cc})^{12} \right\vert^2+ \left\vert (c_t^{cc})^{13} \right\vert^2}\right]^{1/2} \,.
\end{aligned}
\ee

Similar ratios can be defined for hadronic decays. Using the modes $\tau\to h\nu$ and $h\to \mu\nu$, one can build the quantity:
\be
\left(\frac{g_\tau}{g_\mu}\right)_\text{had}=\left[\frac{\mathcal{B}(\tau\to h \nu)}{\mathcal{B}(h\to\mu\bar\nu)} \frac{2 m_h m_\mu^2\tau_h}{(1+\delta R_{\tau/h})m_\tau^3\tau_\tau}\left(\frac{1-m_\mu^2/m_h^2}{1-m_h^2/m_\tau^2}\right)^2 \right]^\frac{1}{2} \,.
\ee
Since the hadronic state $h$ consists of light quarks and must be positively charged, the valence quarks in 
the final state are  $[h]_{\rm val}^-=\bar{u}d_i$, with $d_i=d,s$. 
In this case, we have tree-level contributions to the decay $\tau \to h\nu$ that are parametrised by the Lagrangian
\be
\mathcal{L}(\tau\to h\nu)=-\frac{4G_F}{\sqrt{2}}V_{u d_j }\left[\left(1+\mathcal{C}_{L}^{1j33}\right)(\bar{u} \gamma_\mu P_L d^j)(\bar{\tau}\gamma^\mu P_L \nu_{\tau})+\mathcal{C}_S^{1j33}(\bar{u} P_R d^j)(\bar{\tau} P_L \nu_{\tau})\right]\,,
\ee 
where
\begin{align}
\mathcal{C}_{L}^{1j\alpha\beta}=& \, \frac{2v^2}{\Lambda^2} \, \frac{1}{V_{ud_j}}\left( V_{ud}[\mathcal{C}_{lq}^{(3)}]^{1j\alpha\beta}+V_{us} [\mathcal{C}_{lq}^{(3)}]^{2j\alpha\beta}+V_{ub}[\mathcal{C}_{lq}^{(3)}]^{3j\alpha\beta}\right)  \,, \\
\mathcal{C}_S^{1j33}=&\, \frac{v^2}{\Lambda^2} \frac{1}{V_{ud_j}}\left( V_{ud} [\mathcal{C}_{leqd}]^{1j\alpha\beta}+V_{us} [\mathcal{C}_{leqd}]^{2j\alpha\beta}+V_{ub} [\mathcal{C}_{leqd}]^{3j\alpha\beta}\right)   \,.
\end{align}
The corresponding branching ratio for $\tau\to h\nu$ is
\be
\mathcal{B}(\tau\to h\nu)=\frac{1}{16\pi^2} G_F^2 \tau_h f_h^2 m_\tau^3 \left(1-\frac{m_h^2}{m_\tau^2}\right)^2 \vert V_{ud_j}\vert^2 \left\vert1+\mathcal{C}_{L}^{1j33}+\frac{m_h^2}{m_\tau (m_u+m_{d_j})}\mathcal{C}_S^{1j33}\right\vert^2 \,,
\label{eq:br_tautoh}
\ee
where $\tau_h$ and $f_h$ are the lifetime and decay constant of the hadron $h$. \\

The branching ratio for $h\to \mu\nu$ can be easily extracted by substituting $h\to \mu$ and $\tau\to h$ in \eq{eq:br_tautoh}. The ratios $\left(g_\tau/g_\mu\right)_\text{had}$ when $h=\pi,K$ therefore read
\begin{align}
\left(\frac{g_\tau}{g_\mu}\right)_\pi = & \left( \frac{\left\vert  1+ \mathcal{C}_L^{1133}+\frac{m_\pi^2}{m_\tau (m_u+m_{d})}\mathcal{C}_S^{1133}\right\vert^2}{\left\vert  1+ \mathcal{C}_L^{1122}+\frac{m_\pi^2}{m_\mu (m_u+m_{d})}\mathcal{C}_S^{1122}\right\vert^2} \right)^\frac{1}{2} \,, \\
\left(\frac{g_\tau}{g_\mu}\right)_K = &  \left( \frac{\left\vert  1+ \mathcal{C}_L^{1233}+\frac{m_K^2}{m_\tau (m_u+m_{d})}\mathcal{C}_S^{1233}\right\vert^2}{\left\vert  1+ \mathcal{C}_{L}^{1222}+\frac{m_K^2}{m_\mu (m_u+m_{d})}\mathcal{C}_S^{1222}\right\vert^2} \right)^\frac{1}{2}   \,.
\end{align}
The experimental values for the LFU tests in $W$ couplings defined above are taken from \cite{Amhis:2019ckw} and reported in \Table{tab:LFU_tau}, along with their correlations.

\begin{table}
\begin{center}
\renewcommand{\arraystretch}{1.2} 
\begin{tabular}{c c c c}
\toprule
Obervable & Measurement & Correlation & SM \\
\midrule
 $\left(\frac{g_\tau}{g_\mu}\right)_\text{lept}$    &  $1.0010(15)$  & \multirow{5}{*}{  $\begin{pmatrix}
  \cdot & \cdot & \cdot & \cdot & \cdot \\[6pt]
  0.53  & \cdot & \cdot & \cdot & \cdot \\[6pt]
  -0.49 & 0.48 & \cdot & \cdot & \cdot \\[6pt]
  0.24 & 0.26 & 0.02   & \cdot & \cdot \\[6pt]
  0.11 & 0.10 & -0.01 & 0.06  & \cdot 
  \end{pmatrix}$}& $1.$ \\
$\left(\frac{g_\tau}{g_e}\right)_\text{lept}$      &  $1.0029(15)$  & & $1.$  \\
$\left(\frac{g_\mu}{g_e}\right)_\text{lept}$     & $1.0019(14) $ & & $1.$ \\
$\left(\frac{g_\tau}{g_\mu}\right)_\pi  $    & $0.9961(27)  $& & $1.$ \\
$\left(\frac{g_\tau}{g_\mu}\right)_K  $   & $0.9860(70)$ & & $1.$ \\[5pt]
\toprule
\end{tabular}
\caption{Experimental measurements and correlations for LFU tests in $\tau$ decays \cite{Amhis:2019ckw}.}
\label{tab:LFU_tau}
\end{center}
\end{table}

\subsection{$\tau\to 3 \mu$}

Semileptonic neutral-current operators also generate LFV decays. In particular, the $\ell\to 3\ell^\prime$ decay is described by the following Lagrangian
\be
\mathcal{L}(\ell^\beta\to 3\ell^\alpha)=-\frac{4 G_F}{\sqrt{2}}\left([\mathcal{C}_{lq}^{(1)}]^{33\beta\alpha}-[\mathcal{C}_{lq}^{(3)}]^{33\beta\alpha}\right) c_t^e (\bar{e}^\beta\gamma^\mu P_L e^\alpha)(\bar{e}^\alpha\gamma_\mu P_L e^\alpha)\,,
\ee
where
\be
c_t^e= \frac{v^2}{\Lambda^2}\frac{3 y_t^2}{32\pi^2}\log\frac{\Lambda^2}{m_t^2} \,,
\ee
and only the leading top contribution is taken into account \cite{Feruglio:2017rjo}.

In a $U_1$ model one finds $[\mathcal{C}_{lq}^{(1)}]^{ij\beta\alpha}=[\mathcal{C}_{lq}^{(3)}]^{ij\beta\alpha}$, so that the decays $\ell\to3\ell^\prime$ do not get contributions at one-loop order.

\subsection{$K\to\pi\nu\bar\nu$}

This family of decays probes $s\to d$ transitions. The relevant effective Lagrangian can be written as
\be
\begin{aligned}
\mathcal{L}(s\to d \nu\bar{\nu})= \frac{4 G_F}{\sqrt{2}} V_{td}V^*_{ts} C^\text{SM}_{ds}\frac{\alpha}{2 \pi}& \sum_{\alpha,\beta=1}^3\left[\delta_{\alpha\beta}+\Delta_{sd}\left([\mathcal{C}_{lq}^{(3)}]^{21\alpha\beta}-[\mathcal{C}_{lq}^{(1)}]^{21\alpha\beta}\right)\right](\bar{s}\gamma^\mu P_L d)(\bar\nu^\alpha\gamma_\mu P_L\nu^\beta)\,,
\end{aligned}
\ee
where $C^\text{SM}_{ds}\approx -8.5 \, e^{0.11 \, i}$ \cite{Buchalla:1993wq}, and
\be
\Delta_{sd}=\, \frac{v^2}{\Lambda^2} \frac{\pi}{\alpha_\text{EM}V_{td}V^*_{ts}}\,.
\ee
From \cite{Bordone:2017lsy}, we get:
\be
\mathcal{B}(K^+\to \pi^+\nu\bar\nu)=\frac{1}{3}\sum_{\alpha=1}^3\mathcal{B}(K^+\to \pi^+\nu^\alpha\bar\nu^\alpha)\left\vert1+\Delta_{sd}\left([\mathcal{C}_{lq}^{(3)}]^{21\alpha\beta}-[\mathcal{C}_{lq}^{(1)}]^{21\alpha\beta}\right)\right\vert^2\,, \\
\ee
where LFV effects are neglected. Within the $U_1$ model, these decays do not receive corrections at one loop. Similar considerations apply to $K_L\to\pi^0\nu\bar{\nu}$. 


\section{Scaling of rotation matrices and flavour spurions}

\label{app:scaling}

In this appendix we summarize the scalings of the flavour spurions and the fermion 
rotation matrices from the FN basis to the mass eigenbasis 
for the 11 viable FN charge assignments summarized in table~\ref{tab:viables_sol}.

\subsection{Rotation matrices}

In the FN setup, the rotation matrices for left-handed quark fields are uniquely defined
by the scaling of the CKM matrix,
\begin{align}
 V_{U_L} \sim V_{D_L} &\sim \left(
\begin{array}{ccc}
1 & \lambda & \lambda^3\\
\lambda & 1 & \lambda^2\\
\lambda^3 & \lambda^2 & 1
\end{array}
\right) \,.
\end{align}
For the left-handed leptons (with massless neutrinos), we obtain two 
different possibilities,
\begin{align}
\mbox{scenarios 1x}: & \qquad  V_{E_L} \sim V_{\nu_L} \sim \left(
\begin{array}{ccc}
1 & \lambda & \lambda^3\\
\lambda & 1 & \lambda^2\\
\lambda^3 & \lambda^2 & 1
\end{array}
\right) \sim V_{\rm CKM} \,,  \\
\mbox{scenarios 2x}: & \qquad  V_{E_L} \sim V_{\nu_L} \sim \left(
\begin{array}{ccc}
1 & \lambda^9 & \lambda^7\\
\lambda^9 & 1 & \lambda^2\\
\lambda^7 & \lambda^2 & 1
\end{array}
\right) \,.
\end{align}
For the right-handed down quarks we find 4 different options,
\begin{align}
\mbox{scenarios 1a, 2a}: & \qquad  V_{D_R}  \sim \left(
\begin{array}{ccc}
1 & \lambda^{13} & \lambda^{13}\\
\lambda^{13} & 1 & 1\\
\lambda^{13} & 1 & 1
\end{array}
\right)  \,, \\
\mbox{scenarios 1b, 2b, 2c}: & \qquad  V_{D_R}  \sim \left(
\begin{array}{ccc}
1 & \lambda^{3} & \lambda^{13}\\
\lambda^{3} & 1 & \lambda^{10}\\
\lambda^{13} & \lambda^{10} & 1
\end{array}
\right)  \,, \\
\mbox{scenarios 1c, 2d, 2e}: & \qquad  V_{D_R}  \sim \left(
\begin{array}{ccc}
1 & \lambda^{3} & \lambda^{7}\\
\lambda^{3} & 1 & \lambda^{4}\\
\lambda^{7} & \lambda^{4} & 1
\end{array}
\right)  \,, \\
\mbox{scenarios 1d, 1e, 2f}: & \qquad  V_{D_R}  \sim \left(
\begin{array}{ccc}
1 & \lambda^{1} & \lambda^{1}\\
\lambda^{1} & 1 & 1\\
\lambda^{1} & 1 & 1
\end{array}
\right)  \,.
\end{align}
Finally, the rotation matrices for right-handed charged leptons are scaling according 
to 

\begin{align}
\mbox{scenarios 1a, 1d}: & \qquad  V_{E_R}  \sim \left(
\begin{array}{ccc}
1 & \lambda^{15} & \lambda^{9}\\
\lambda^{15} & 1 & \lambda^{6}\\
\lambda^{9} & \lambda^{6} & 1
\end{array}
\right)  \,, \\
\mbox{scenario 1b}: & \qquad  V_{E_R}  \sim \left(
\begin{array}{ccc}
1 & \lambda^{5} & \lambda^{9}\\
\lambda^{5} & 1 & \lambda^{4}\\
\lambda^{9} & \lambda^{4} & 1
\end{array}
\right)  \,, \\
\mbox{scenario 1c}: & \qquad  V_{E_R}  \sim \left(
\begin{array}{ccc}
1 & \lambda^{5} & \lambda^{15}\\
\lambda^{5} & 1 & \lambda^{10}\\
\lambda^{15} & \lambda^{10} & 1
\end{array}
\right)  \,, \\
\mbox{scenario 1e}: & \qquad  V_{E_R}  \sim \left(
\begin{array}{ccc}
1 & \lambda^{3} & \lambda^{9}\\
\lambda^{3} & 1 & \lambda^{6}\\
\lambda^{9} & \lambda^{6} & 1
\end{array}
\right)  \,, \\ 
\mbox{scenarios 2a, 2f}: & \qquad  V_{E_R}  \sim \left(
\begin{array}{ccc}
1 & \lambda^{13} & \lambda^{19}\\
\lambda^{13} & 1 & \lambda^{6}\\
\lambda^{19} & \lambda^{6} & 1
\end{array}
\right)  \,, \\
\mbox{scenario 2b}: & \qquad  V_{E_R}  \sim \left(
\begin{array}{ccc}
1 & \lambda^{5} & \lambda^{1}\\
\lambda^{5} & 1 & \lambda^{4}\\
\lambda^{1} & \lambda^{4} & 1
\end{array}
\right)  \,, \\
\mbox{scenario 2c}: & \qquad  V_{E_R}  \sim \left(
\begin{array}{ccc}
1 & \lambda^{23} & \lambda^{19}\\
\lambda^{23} & 1 & \lambda^{4}\\
\lambda^{19} & \lambda^{4} & 1
\end{array}
\right)  \,, \\
\mbox{scenario 2d}: & \qquad  V_{E_R}  \sim \left(
\begin{array}{ccc}
1 & \lambda^{5} & \lambda^{5}\\
\lambda^{5} & 1 & \lambda^{10}\\
\lambda^{5} & \lambda^{10} & 1
\end{array}
\right)  \,, \\ 
\mbox{scenario 2e}: & \qquad  V_{E_R}  \sim \left(
\begin{array}{ccc}
1 & \lambda^{23} & \lambda^{13}\\
\lambda^{23} & 1 & \lambda^{10}\\
\lambda^{13} & \lambda^{10} & 1
\end{array}
\right)  \,.
\end{align}

\subsection{Flavour spurions}

For the flavour spurions we obtain
\begin{align}
 \mbox{scenarios 1x:} & \qquad 
 \Delta_{QL} \sim \left(
\begin{array}{ccc}
\lambda^5 & \lambda^{4} & \lambda^{2}\\
\lambda^{4} & \lambda^3 & \lambda^{1}\\
\lambda^{2} & \lambda^{1} & \lambda^1
\end{array}
\right) \,,  \\
 \mbox{scenarios 2x:} & \qquad 
 \Delta_{QL} \sim \left(
\begin{array}{ccc}
\lambda^5 & \lambda^{4} & \lambda^{2}\\
\lambda^{6} & \lambda^3 & \lambda^{1}\\
\lambda^{8} & \lambda^{1} & \lambda^1
\end{array}
\right) \,, 
\end{align}
together with
\begin{align}
 \mbox{scenarios 1a:} & \qquad 
 \Delta_{DE} \sim \left(
\begin{array}{ccc}
\lambda^{21} & \lambda^{6} & \lambda^{12}\\
\lambda^{8} & \lambda^7 & \lambda^{1}\\
\lambda^{8} & \lambda^{7} & \lambda^1
\end{array}
\right) \,, \\
 \mbox{scenarios 1b:} & \qquad 
 \Delta_{DE} \sim \left(
\begin{array}{ccc}
\lambda^{21} & \lambda^{16} & \lambda^{12}\\
\lambda^{18} & \lambda^{13} & \lambda^{9}\\
\lambda^{8} & \lambda^{3} & \lambda^1
\end{array}
\right) \,, \\
 \mbox{scenarios 1c:} & \qquad 
 \Delta_{DE} \sim \left(
\begin{array}{ccc}
\lambda^{21} & \lambda^{16} & \lambda^{6}\\
\lambda^{18} & \lambda^{13} & \lambda^{3}\\
\lambda^{14} & \lambda^{9} & \lambda^1
\end{array}
\right) \,, \\
 \mbox{scenarios 1d:} & \qquad 
 \Delta_{DE} \sim \left(
\begin{array}{ccc}
\lambda^{7} & \lambda^{8} & \lambda^{2}\\
\lambda^{8} & \lambda^7 & \lambda^{1}\\
\lambda^{8} & \lambda^{7} & \lambda^1
\end{array}
\right) \,, \\ 
 \mbox{scenarios 1e:} & \qquad 
 \Delta_{DE} \sim \left(
\begin{array}{ccc}
\lambda^{11} & \lambda^{8} & \lambda^{2}\\
\lambda^{10} & \lambda^7 & \lambda^{1}\\
\lambda^{10} & \lambda^{7} & \lambda^1
\end{array}
\right) \,, 
\end{align}
and 
\begin{align}
 \mbox{scenarios 2a:} & \qquad 
 \Delta_{DE} \sim \left(
\begin{array}{ccc}
\lambda^{7} & \lambda^{6} & \lambda^{12}\\
\lambda^{20} & \lambda^7 & \lambda^{1}\\
\lambda^{20} & \lambda^{7} & \lambda^1
\end{array}
\right) \,, \\
 \mbox{scenarios 2b:} & \qquad 
 \Delta_{DE} \sim \left(
\begin{array}{ccc}
\lambda^{11} & \lambda^{16} & \lambda^{12}\\
\lambda^{8} & \lambda^{13} & \lambda^{9}\\
\lambda^{2} & \lambda^{3} & \lambda^1
\end{array}
\right) \,, \\
 \mbox{scenarios 2c:} & \qquad 
 \Delta_{DE} \sim \left(
\begin{array}{ccc}
\lambda^{7} & \lambda^{16} & \lambda^{12}\\
\lambda^{10} & \lambda^{13} & \lambda^{9}\\
\lambda^{20} & \lambda^{3} & \lambda^1
\end{array}
\right) \,, \\
 \mbox{scenarios 2d:} & \qquad 
 \Delta_{DE} \sim \left(
\begin{array}{ccc}
\lambda^{11} & \lambda^{16} & \lambda^{6}\\
\lambda^{8} & \lambda^{13} & \lambda^{3}\\
\lambda^{4} & \lambda^{9} & \lambda^1
\end{array}
\right) \,,  \\
 \mbox{scenarios 2e:} & \qquad 
 \Delta_{DE} \sim \left(
\begin{array}{ccc}
\lambda^{7} & \lambda^{16} & \lambda^{6}\\
\lambda^{10} & \lambda^{13} & \lambda^{3}\\
\lambda^{14} & \lambda^{9} & \lambda^1
\end{array}
\right) \,, \\
 \mbox{scenarios 2f:} & \qquad 
 \Delta_{DE} \sim \left(
\begin{array}{ccc}
\lambda^{21} & \lambda^{8} & \lambda^{2}\\
\lambda^{20} & \lambda^{7} & \lambda^{1}\\
\lambda^{20} & \lambda^{7} & \lambda^1
\end{array}
\right) \,.
\end{align}

\bibliographystyle{utphys.bst}

{\footnotesize
\bibliography{references}
}

\end{document}